\documentclass[12pt]{article}
\usepackage{amsfonts,amsmath}
 \usepackage{amssymb}
\usepackage{mathrsfs}

\usepackage[hypertex] {hyperref}

\makeatletter
\@addtoreset{equation}{section}
\makeatother
\def\theequation{\thesection.\arabic{equation}}
\newcommand{\eq}{\eqref}
 \newcommand{\LLL}{\mathcal{J}{\,}}

\newcommand{\pss}{\mathbf{s}}

\newcommand{\hf}{\mathbf{n}}

\newcommand{\dt}{{\vartriangle}}
\newcommand{\pt}{{\bar\vartriangle}}
\newcommand{\te}{ {\theta}}

\newcommand{\bgt}{{\bar{\gt}}}
 \newcommand{\bt}{{\bar{t}}}
 \newcommand{\dis}{\displaystyle}

\newcommand{\stt}{{*}}
\newcommand{\sstt}{{*}}

 \newcommand{\bs}{{\bar s}}

\newcommand{\dr}{{\rm d}}
\newcommand{\be}{ \begin{equation}}

\newcommand{\etc}{{\it etc}}
\newcommand{\slv}{{}^v{}\mathfrak{sl}_2}
\newcommand{\slh}{{}^h{}\mathfrak{sl}_2}

\newcommand{\bp}{\bar{\partial}}

\newcommand{\GGG}{\mathit{\Gamma}}
\newcommand{\GGS}{\mathit{\Gamma}}

\newcommand{\bm}{\bar{m}}

\newcommand{\NNN}{{\mathcal N}}

\newcommand{\III} {\Lambda}%

\newcommand{\ee}{\end{equation}}
\newcommand{\PPP}{J}

\newcommand{\bee}{\begin{eqnarray}}
\newcommand{\beee}{\begin{array}}
\newcommand{\eee}{\end{eqnarray}}
\newcommand{\eeee}{\end{array}}

\newcommand{\gt}{\tau}

\newcommand{\gr}{\rho}
\newcommand{\ga}{\alpha}
\newcommand{\gb}{\beta}
\newcommand{\gga}{\gamma}

\newcommand{\dgb}{{\dot \gb}}
\newcommand{\dga}{{\dot \ga}}

\newcommand{\pa}{{\dot{\ga}}}
\newcommand{\pb}{{\dot{\gb}}}
\newcommand{\pga}{{\dot{\gamma}}}

  \newcommand{\pd}{{\dot{\delta}}}

 \newcommand{\F}{{\cal F}}

\newcommand{\Hh}{{\mathscr H}} 
\newcommand{\Hhh}{{ \cal{H}}}
\newcommand{\HhhH}{{ \cal{G}}}

\newcommand{\rhs}{{\it r.h.s.} }

\newcommand{\ie}{{\it i.e.,} }
\newcommand{\ls}{\!\!\!\!\!\!}

\newcommand{\gd}{\delta}

\newcommand{\gl}{\lambda}

\newcommand{\gvep}{\varepsilon}

\newcommand{\gs}{\sigma}

\newcommand{\go}{\omega}

\newcommand{\by}{{\bar{y}}}

\newcommand{\q}{\,,\qquad}
  \newcommand{\nn}{\nonumber}

\newcommand{\half}{\frac{1}{2}}

\newcommand{\p}{\partial}
\newcommand{\D}{{\cal D}}
\newcommand{\f}{\frac}
\newcommand{\A}{{\cal A}}

\newcommand{\CCC}{{C}}

\newsavebox{\ver}
\newsavebox{\verp}
\newsavebox{\gorp}
\newsavebox{\toch}

\begin{document}
\begin{flushright}

{\small FIAN/TD/03-17}
\end{flushright}
\vspace{1.7 cm}

\begin{center}
{\large\bf Current Interactions from
the One-Form Sector of\\
  Nonlinear Higher-Spin Equations
}

\vspace{0.7 cm}
O.A.~Gelfond$^{1,2}$ and M.A.~Vasiliev$^2$ \vglue 0.3  true cm

\vspace{0.7 cm}
${}^1$ Federal Scientific Centre Scientific Research Institute of System Analysis
of  Russian Academy of Science,\\
Nakhimovsky prospect 36-1, 117218, Moscow, Russia

 \vglue 0.3  true cm

 ${}^2$I.E.Tamm Department of Theoretical Physics, Lebedev Physical Institute,\\
Leninsky prospect 53, 119991, Moscow, Russia
 \vskip1.5cm
\date{date}
\end{center}

\vspace{0.4 cm}

 \begin{abstract}
\noindent The form of higher-spin current interactions in the sector of one-forms is
derived  from the  nonlinear higher-spin equations in $AdS_4$.
Quadratic corrections to higher-spin
equations are shown to be independent of the phase of  the parameter  $\eta =\exp i\varphi$
  in the full nonlinear higher-spin equations.
  The current deformation  resulting from the nonlinear higher-spin equations
 is represented in the canonical form with the minimal number of space-time
 derivatives. The non-zero spin-dependent coupling constants of the resulting currents are
   determined in terms of the higher-spin coupling constant $\eta\bar\eta$.
 Our results  confirm the
 conjecture that (anti-)self-dual  nonlinear higher-spin equations result
 from the full system at ($\eta=0$) $\bar \eta=0$.

\end{abstract}

\newpage
\tableofcontents
\newpage

\section{Introduction}\label{Generalities}

Though nonlinear field equations for massless higher-spin (HS) fields
in various dimensions are available for a long time
\cite{Vasiliev:1990en,more,prok,Vasiliev:2003ev} their structure beyond
the linearized level still is not fully understood.
As shown in
\cite{Bengtsson:1983pd,Bengtsson:1983pg,Berends:1984ws,Berends:1984rq}, HS
interactions consistent with HS gauge symmetries contain higher derivatives
though no higher derivatives appear at the quadratic level in a
maximally symmetric background geometry \cite{Frhs,Frfhs}. Along with the fact,
that a consistent HS theory containing a propagating field of any spin $s>2$ should
necessarily contain an infinite tower of HS fields of infinitely increasing spins
originally indicated by the analysis of HS symmetries in
\cite{Berends:1984ws,Berends:1984rq} and later shown to follow from
the structure of HS symmetry algebras \cite{Fradkin:1986ka,OP1}, this implies that
any HS gauge theory is somehow nonlocal.

Appearance of higher derivatives in interactions demands a
dimensionful coupling constant $\rho$ which was identified in
 \cite{Fradkin:1987ks,Fradkin:1986qy} with the radius of the background
$(A)dS$ space. Resulting higher-derivative
vertices allow no meaningful flat limit in agreement with numerous
no-go statements ruling out nontrivial interactions of massless HS
fields in Minkowski space \cite{Coleman:1967ad,Aragone:1979hx} (see
 \cite{Bekaert:2010hw} for more detail and references).
Geometric origin of the dimensionful parameter $\rho$
has an important consequence that any HS gauge theory  with
unbroken HS symmetries does not allow a  parametric low-energy analysis
with respect to a large scale parameter like Plank energy or $\alpha'$ because
the rescaled covariant derivatives $\D=\rho D$ in the
expansion in powers of derivatives
\be
\label{exp}
 \sum_{n,m=0}^\infty a_{nm} \D^n \phi \D^m \phi+\ldots\,
\ee
 cannot be treated as small
since, being non-commutative in the background $AdS$ space-time of curvature $\rho^{-2}$,
 they have commutator of order one, $[\D\,,\D] \sim 1$.
As a result,  all terms in (\ref{exp}) may give comparable contributions. Whether
expansion (\ref{exp}) is local or not   depends  on the behavior of the
coefficients $a_{nm}$ at $n,m\to \infty$. If at most a finite number of coefficients
$a_{n,m}$ is nonzero,  field redefinition (\ref{exp}) is genuinely local.

Importance of the proper definition of locality was originally stressed in \cite{prok}
where it was shown that by a field redefinition involving expansion of the form (\ref{exp})
it is possible to get rid of the currents from $3d$  HS field equations. Recently
this issue was reconsidered  in \cite{Vasiliev:2015wma}, where
a proposal was put forward on the part of the problem associated with the
exponential factors resulting from so-called inner Klein operators while the
structure of the pre-exponential factors was only partially determined.
The issue of locality was also analyzed in
 \cite{Boulanger:2015ova,Bekaert:2015tva,Skvortsov:2015lja}. Focusing on the
 lowest-order current-type interactions the authors of
  \cite{Boulanger:2015ova,Skvortsov:2015lja} failed to find the appropriate scheme
  of the analysis of nonlinear HS equations,
  arriving at the conclusion that it may be hard to distinguish between local and nonlocal
  frames in the setup  of \cite{more}.

  On the other hand, in \cite{Vasiliev:2016xui} a simple field redefinition was
found that brings the quadratic corrections to the field equations in the sector
of zero-forms to the canonical form of local current interactions  found
originally in \cite{Gelfond:2010pm}. The field redefinition of \cite{Vasiliev:2016xui}
had a simple form, bringing
the HS  equations to the local form in the lowest order (in what follows
this field redefinition as well as its extension to the one-form sector
considered in this paper will be referred to as proper). As explained in
 \cite{Vasiliev:2016xui}, this field redefinition is unique within the natural
Ansatz  expressing the separation of variables between the sectors of left and right spinors in the theory.
Let us stress that
 the nonlocal field
redefinitions applied in this paper as well as in \cite{Vasiliev:2016xui} is not a principle issue but rather
a technical tool relating the proper local formulation obtained   with the originally known (improper) nonlocal one.

One of the surprising outputs
of the analysis of \cite{Vasiliev:2016xui} was that nonlinear HS
equations properly reproduce
usual current interactions of higher spins with the coupling constant
independent of the phase parameter $\varphi$ distinguishing between inequivalent HS equations.
The aim of this paper is to extend the results of \cite{Vasiliev:2016xui} to the sector of
equations on HS one-form gauge potentials bringing their right-hand-sides to the standard local
 current form.
 We will show that indeed there exists a choice of field variables leading to the
 proper  result and compute the coupling coefficients in front of different currents on the right-hand-side of HS equations.  Let us stress that this choice of variables is uniquely determined by that
 of \cite{Vasiliev:2016xui} up to the gauge transformations of HS gauge fields and local field redefinitions.
 Again, as in the zero-form sector, the resulting coupling
constants turn out to be independent of the phase  $\varphi$.

The obtained results provide a basis for the analysis of locality in HS theory
along the lines of   \cite{1707.03735} where it is shown in particular that
the field redefinition found in  \cite{Vasiliev:2016xui} is the only proper one, hence leading to unambiguous results for the HS current coupling constants. Moreover, as stressed in \cite{1707.03735},  the necessity of the nonlocal field
redefinitions found in \cite{Vasiliev:2016xui} is a consequence of
the improper  choice of resolution operator in the process of solving the nonlinear  HS equations in the sector of auxiliary $Z$-variables while an alternative choice of the
``local resolution operator" leads directly to the correct local result of \cite{Vasiliev:2016xui}. The results of this paper are anticipated to shed light on
the form of the  local resolution operator in the one-form sector as well.

There are two important consequences of the independence of the HS coupling of
the phase $\varphi$ which is the phase of the complex conjugated  parameters
\be
\eta = |\eta| \exp i\varphi\q \bar \eta = |\eta| \exp -i\varphi\,
\ee
in the nonlinear HS equations of \cite{more}. One is that in \cite{more}
it was conjectured that the $4d$ HS theory with ($\eta $) $\bar \eta=0$ describes an
(anti-)self-dual HS theory. The conclusion reached in \cite{Vasiliev:2016xui} and in this
paper that the current interaction terms are proportional to $\eta\bar\eta$ implies
in particular that all of them do not contribute to the purely self-dual sector, which
result is anticipated because, as is well known, no nontrivial amplitude can be constructed
in the purely self-dual sector. Hence, our results provide a nontrivial support for the
conjecture of \cite{more} on the  (anti-)self-dual HS theory. The study of the latter theory which,
as we show, needs a  special choice of dynamical variables identified in this
paper to the second order, is itself  very interesting.

Another interesting direction is the HS holography (see e.g.
\cite{Sundborg:2000wp}-\cite{Giombi:2012ms}). The nontrivial question here is that
the parity non-invariant HS theories with general phase $\varphi$ were conjectured in
\cite{Aharony:2011jz,Giombi:2011kc} to be dual to certain parity breaking Chern-Simons
boundary theories. Though this conjecture seemingly contradicts to the conclusion that
the HS cubic vertices are independent of $\varphi$, in
\cite{Vasiliev:2016xui} it was argued that this is not the case and there is precise
matching of the obtained $\varphi$-independent HS vertices with the structure of the
boundary three-point function found in \cite{Maldacena:2012sf}. The analysis of
\cite{Vasiliev:2016xui} was performed in terms of certain boundary conditions on the
HS connections, which may look unusual from the perspective of the standard approach.
In \cite{Didenko:2017lsn}
     this question is reconsidered  in a more standard way with the
same final result.

The rest of the paper is organized as follows.
 In Section \ref{HSequat}  we recall  relevant features of the
the full nonlinear HS equations in $AdS_4$ and  their perturbative
analysis. In Section \ref{Second order}
 the  quadratic corrections are discussed. General structure of current interactions is
 recalled  in Section \ref{Currinter}.
  Section \ref{Quadraticc}
recalls the analysis of quadratic corrections in the zero-form sector, including the field redefinition
of \cite{Vasiliev:2016xui} bringing the
quadratic corrections in this sector to the local form.
  Section \ref{Main results} summarizes  the main results.
 In Section \ref{nonlincorone}
 quadratic corrections in the one-form sector are considered. Specifically,
 the quadratic corrections   from nonlinear equations are found in Section
 \ref{nonlinetaeta}.
  It is shown   that,  modulo exact forms, the sum of quadratic corrections resulting
  from the zero-form redefinition with those coming from the nonlinear
 equations contains only $\eta\bar\eta$ terms.  The appropriate  field redefinition,
 that brings the
$\eta\bar\eta$-proportional quadratic corrections in the one-form sector
to the local form, is found in Section  \ref{etabaretaloc}.
 Flat limit rescalings   are recalled in Section \ref{Flat limit }.
 The local field redefinition bringing the  local
  quadratic corrections in the one-form sector to the form allowing
  flat limit is also presented here.
In Section \ref{contribution}, it is shown in detail
 how the  resulting second-order corrections of unfolded equations
 bring currents to the right-hand sides of
 the Fronsdal-like dynamical equations for massless fields.
Conclusions and perspectives are briefly
discussed in Section \ref{conc}. Some  useful formulae are collected in
Appendix A. Appendix B   presents details of the derivation of field redefinitions
 in the one-form sector.

\section{Preliminaries}
\subsection{Higher-spin equations in $AdS_4$}
\label{HSequat}
$4d$ nonlinear HS equations have the form \cite{more}
\begin{eqnarray}
 &  & \mathrm{d} W+W*\wedge W=-i\theta_{\alpha}\wedge\theta^{\alpha}
 \left(1+\eta B*\varkappa* k\right)-i\bar{\theta}_{\dot{\alpha}}\wedge
 \bar{\theta}^{\dot{\alpha}}\left(1+\bar{\eta}B*\bar{\varkappa}*\bar{k}\right),
 \label{eq:HS_1}\qquad\\
 &  & \mathrm{d} B+W*B-B*W=0.\label{eq:HS_2}\qquad
\end{eqnarray}
Here $\mathrm{d} =dx^{\underline{m}}\tfrac{\partial}{\partial x^{\underline{m}}}$
is the space-time de Rham differential (onwards wedge symbol is omitted).
$B(Z;Y;K|x)$ and $W(Z;Y;K|x)$ are fields of the theory which depend both on space-time
coordinates $x$ and on spinorial variables $Y^{A}=\left(y^{\alpha},\bar{y}^{\dot{\alpha}}\right)$
and $Z^{A}=\left(z^{\alpha},\bar{z}^{\dot{\alpha}}\right)$ where
the spinor indices $\alpha$ and $\dot{\alpha}$ take two values.
The noncommutative star product $*$ acts on functions of $Y$ and $Z$
\begin{equation}
\left(f*g\right)\left(Z;Y\right)=\int d^{4}Ud^{4}Vf\left(Z+U;Y+U\right)g\left(Z-V;Y+V\right)\mathrm{e}^{iU_{A}V^{A}},\label{eq:star_product}
\end{equation}
 where $U_{A}V^{A}=U^{A}V^{B}\epsilon_{AB}=u^{\alpha}v^{\beta}\epsilon_{\alpha\beta}+\bar{u}^{\dot{\alpha}}\bar{v}^{\dot{\beta}}\epsilon_{\dot{\alpha}\dot{\beta}}$
and $\epsilon_{AB}$ is the $sp\left(4\right)$-invariant symplectic form built from
the $sp\left(2\right)$-forms $\epsilon_{\alpha\beta}$, $\epsilon_{\dot{\alpha}\dot{\beta}}$.
Integration measure $d^{4}Ud^{4}V$ in \eqref{eq:star_product} is normalized so that
 $f*1=f$, \ie the factor of $\f{1}{(2\pi)^4}$ is implicit.

Inner Klein operators $\varkappa$ and $\bar{\varkappa}$
\begin{equation}
\varkappa:=\exp\left(iz_{\alpha}y^{\alpha}\right),\qquad\bar{\varkappa}:=
\exp\left(i\bar{z}_{\dot{\alpha}}\bar{y}^{\dot{\alpha}}\right)\,\label{Klein}
\end{equation}
have the  properties
\begin{eqnarray}
 &  & \varkappa*\varkappa=1,\qquad\varkappa*f\left(z^{\alpha},y^{\alpha}\right)=f\left(-z^{\alpha},-y^{\alpha}\right)*\varkappa,
  \\
 &  & f\left(y,z\right)*\varkappa=f\left(-z,-y\right)e^{iz_{\alpha}y^{\alpha}},\label{eq:f*kappa}
\end{eqnarray}
and analogously for $\bar{\varkappa}$.

\textbf{$B$ }is a zero-form, while $W$ is a one-form in the space-time
differential $dx^{\underline{m}}$ and  auxiliary differential
$\theta^{A}$ dual to $Z^{A}$
\be
W(Z;Y;K|x)=dx^{\underline{m}} W_{\underline{m}}(Z;Y;K|x) + \theta^A S_A\,.
\ee
 All differentials anticommute
\begin{equation}
\left\{ dx^{\underline{m}},dx^{\underline{n}}\right\} =\left\{ dx^{\underline{m}},\theta^{A}\right\} =\left\{ \theta^{A},\theta^{B}\right\} =0.
\end{equation}

In addition to the inner Klein operators of the star-product algebra
there is also a pair of outer Klein operators $k$ and $\bar{k}$
which have similar properties
\begin{equation}\label{EQdef}
k*k=1,\qquad k*f\left(z^{\alpha};y^{\alpha};\theta^{\alpha}\right)=
f\left(-z^{\alpha};-y^{\alpha};-\theta^{\alpha}\right)*k\,.
\end{equation}
However, being anticommutative with (anti)holomorhic $\theta$ differentials, they  admit no star-product
algebra  realization. The  fields $W(Z;Y;K|x)$ and $B(Z;Y;K|x)$ depend on the exterior Klein
operators. (Relations (\ref{EQdef}) provide a definition of the $*$-product for $k$ and $\bar k$.)
The sector of physical fields is represented by $B(Z;Y;K|x)$ linear in $k$
or $\bar{k}$, while $W(Z;Y;K|x)$ instead depends on $k\bar{k}$.

For topological
sector this is the other way around. The latter is truncated away in this paper.
$\eta$ in \eqref{eq:HS_1} is a free complex parameter
which can be normalized to be unimodular $|\eta|=1$ hence representing
the phase factor freedom $\eta=\exp i{\varphi}$.

Background $AdS_4$ space of radius $\lambda^{-1}=\rho$
is described by a flat $sp(4)$
connection $w=(\go_L{}^{\alpha \gb},\overline{\go{}}^{\pa\pb}_L,h^{\ga\pb})$
containing Lorentz connection
$\go_L{}^{\alpha \gb} $, $\overline{\go{}}^{\pa\pb}_L$ and
vierbein  $h^{\ga\pb}$ that obey the flatness conditions
\be
\label{adsfl}
R^{\ga\gb}=0\,,\quad \overline{R}^{\pa\pb}=0\,,
\quad R^{\ga\pa}=0\,,
\ee
where
\be
\label{nR}
R^{\alpha \gb}=\dr\go_L{}^{\alpha \gb} +\go_L{}^{\alpha}{}_\gamma
 \go_L{}^{\gb}{}^{\gamma} -\lambda^2\, H^{\alpha \gb}
\q\overline{R}^{{\pa} {\pb}}
=\dr\overline{\go}_L{}^{{\pa}
{\pb}} +\overline{\go}_L{}^{{\pa}}{}_{\dot{\gamma}}
 \overline{\go}_L{}^{{\pb}}{}^{ \dot{\gga}} -\lambda^2\,
 \bar{H}^{{\pa\pb}}\,,
\ee
\begin{equation}
\label{nr}
R^{\alpha {\pb}} =\dr h^{\alpha{\pb}} +\go_L{}^\alpha{}_\gamma
h^{\gamma}{}^{\pb} +\overline{\go}_L{}^{{\pb}}{}_{\pd}
 h^{\alpha}{}^{\pd}\,,
\end{equation}
 where  \be\label{znakH}
H^{\ga\gb}=h^\ga{}^\pa h^\gb{}_\pa\q \bar{H}^{\pa\pb}=h^\pa{}^\ga h^\pb{}_\ga
\q h^\ga{}^\pa h^\gb{}^\pb=-\half\gvep^{\ga\gb}\bar{H}^{\pa\pb}-
\half\gvep^{\pa\pb} {H}^{\ga\gb}\,.
\ee


 In the first nontrivial order
\be
\label{B1C}
B_1(Z;Y;K|x) = C(Y;K|x)\,.
\ee
Thus, nonlinear HS equations (\ref{eq:HS_1}), (\ref{eq:HS_2}) give rise to
 the doubled set of massless fields
\be\label{Csumkbark}
C(Y;K|x)= C^{1,0}(Y|x) k + C^{0,1}(Y|x) \bar k\,.
\ee

According to \cite{more}, nontrivial part of  linearized equations
 \eqref{eq:HS_1}, \eqref{eq:HS_2} has the form of so-called Central-on-shell theorem originally
 found  in \cite{Ann}
\bee
\label{CON1k}
    &&  \D_{ad} \go(y,\by ;K|x) =L(C)\q\\
\label{CON2k}
\,&&    \D_{tw} C (y,\by ;K| x) =0\,,
\eee
where the term  $L(C)$ linear in $C$ reads as
\be\label{LotC}    L(C)=\frac{i\gl}{4} \left ( \eta \bar{H}^{\dga\pb}\! {\bp_{\dga}\bp_{\dgb} }
{C}(0,\bar y;K| x)*k +\bar \eta H^{\ga\gb}\! \p_{\ga}\p_{\gb}\,{C}(y,0;K| x)*\bar k \right )  \q  \ee
spin-$s$ one-form  $\go$ is
 \be\label{spinsgo}\go(y,{\bar{y}};K|x)=\f{1}{2i}\sum_{m,n\ge0}\f{1}{m!n!}
\omega{}_{\alpha_1\ldots\alpha_n}{}_,{}_{{\dot \gb}_1
\ldots{\dot \gb}_m}(K|x)y^{\alpha_1}\dots y^{\alpha_n} \by^{\dot\gb_1}\dots
\by^{\dot\gb_m}\,
\ee
with $n+m=2(s-1)$ (for   $ s \ge 1$)\,, spin-$s$ zero-form
$C(y,{\bar{y}};K|x)$ (\ref{Csumkbark})
 \be\label{spinsC}C^{ij}(y,{\bar{y}} |x)=\f{1}{2i}\sum_{m,n\ge0}\f{1}{m!n!}C^{ij}{}_{\alpha_1\ldots\alpha_n}{}_,{}_{{\dot\gb}_1
\ldots{\dot\gb}_m}(x) y^{\alpha_1}\dots y^{\alpha_n}
\by{}^{\dot\gb_1}\dots \by{}^{\dot\gb_m} , \ee
 has $|n-m|=2s$, and
 \be
\label{Dad}
  \D_{ad}\omega(y,{\bar{y}};K| x) :=
D^L \omega (y,\bar{y};K| x) +
\lambda h^{\ga\pb}\Big (y_\ga \bp_\pb
+\p_\ga\bar{y}_\pb\Big )
\omega (y,\bar{y} ;K| x) \,,
\ee
\be
\label{tw}
 \D_{tw} C(y,{\bar{y}};K| x) :=
D^L C (y,{\bar{y}};K| x) -{i}\lambda h^{\ga\pb}
\Big (y_\ga \bar{y}_\pb -\p_\ga\bp_\pb
\Big ) C (y,{\bar{y}};K| x)\,,
\ee
\be
\label{dlor}
D^L f (y,{\bar{y}};K| x) :=
\dr f (y,{\bar{y}};K| x) +
\Big (\go_L^{\ga\gb}y_\ga \p_\gb
+
\overline{\go}_L^{\pa\pb}\bar{y}_\pa\bp_\pb
\Big )
f (y,{\bar{y}};K| x)\,\,
,\ee \be\nn\p {}_\ga =
 \frac{\partial }{\partial  y {}^\ga}\q
\bp {}_\pa =
 \frac{\partial }{
\partial \bar{y} {}^\pa}\q \dr  = dx^{\underline{n}}\f{\p}{\p x^{\underline{n}}}\,. \ee
Equations \eq{CON1k}, \eq{CON2k} are equivalent to usual massless Fronsdal equations \cite{Frhs,Frfhs}
supplemented by an infinite set of auxiliary fields and constraints. The Fronsdal fields are
contained in the frame-like fields $\omega{}_{\alpha_1\ldots\alpha_n}{}_,{}_{{\dot\gb}_1
\ldots{\dot\gb}_m}(x)$ with $n=m$ for bosons and $|n-m|=1$ for fermions.

 Our goal is to find the second-order corrections to Central-on-shell theorem resulting from
 nonlinear equations  \eqref{eq:HS_1}, \eqref{eq:HS_2}.
To simplify formulae in the sequel we set  $\gl=1$.

\subsection{Perturbative analysis}
\label{Second order}
To start a perturbative expansion one has to fix some vacuum solution
to \eqref{eq:HS_1}, \eqref{eq:HS_2}. Eq.~\eqref{eq:HS_2} can
be solved by setting the vacuum value of ${B}$ to zero
\begin{equation}\nn
{B}_{0}=0.
\end{equation}
A natural vacuum solution for \eqref{eq:HS_2}
is
\begin{equation}
W_{0}=\phi_{AdS}+Z_{A}\theta^{A},\label{eq:W_0}
\end{equation}
where $\phi_{AdS}$ is the space-time $sp\left(4\right)$ connection  one-form
describing the $AdS_{4}$ background
\be \label{fads}\phi_{AdS}=-\dfrac{i}{4}\Big(\omega_L^{\alpha\beta}y_{\alpha}y_{\beta}+
 \bar{\omega}_L^{\dot{\alpha}\dot{\beta}}\bar{y}_{\dot{\alpha}}\bar{y}_{\dot{\beta}}+
 2h^{\alpha\dot{\beta}}y_{\alpha}\bar{y}_{\dot{\beta}}\Big)
\ee
with Eq.~(\ref{adsfl}) taking the form
\be
 \mathrm{d} \phi_{AdS}+\phi_{AdS}*\phi_{AdS}=0\,.\label{eq:dw_ww}
\ee

To the second order, nonlinear equations
 \eqref{eq:HS_1}, \eqref{eq:HS_2}  have the form
\bee\label{order2}
   &&\Delta_{ad}W_{2}+W_{1}*W_{1}=-i\eta B_{2}* k* \varkappa*  \theta_{\alpha}\theta^{\alpha}
   -i\bar{\eta}B_{2}*\bar k * \bar\varkappa* \bar{\theta}_{\pa}\bar{\theta}^{\pa}\,,\\\nn
 && \Delta_{tw}B_{2}+\left[W_{1},B_{1}\right]_{*}=0\,,
\eee where $B_{2}$ and $W_{2}$ are second-order fields, \bee
\Delta_{ad}f&:=&\mathrm{d} f+\left[\phi_{AdS},f\right]_{*}-2i\mathrm{d}_{Z}f  \label{deltas}
\\\nn\Delta_{tw}f &:=&\mathrm{d} f-\frac{i}{2}\left[\omega^{AB}Y_{A}Y_{B},f\right]_{*}-
\frac{i}{2}\left\{ h^{AB}Y_{A}Y_{B},f\right\} _{*}-2i\mathrm{d}_{Z}f\q
\\ \nn \mathrm{d}_{Z}&:=&\theta^{A}\tfrac{\partial}{\partial Z^{A}}\,.\eee
 More precisely, the expansion should be interpreted as a filtration, \ie $W_2$ and  $B_2$ contain
 all terms up to order two rather than just the second-order part.

Equations on the $Z$-independent part (cohomology) of the $B_{2}$-field, $C(Y;K|x)$ \eq{Csumkbark}, are  \cite{Didenko:2015cwv} \,:
\be\label{C1}
\mathcal{D}_{tw}C=-  \Hh_{tw}\left(W_{1}*C-C*\pi\left(W_{1}\right)\right),
\ee where $\pi(y,\bar{y})=(-y,\bar{y})$ and 
\be \label{W1}W_{1}=\omega\left(Y\right)-i\Delta_{ad}^{*}\left(\eta
C* k *  \varkappa*  {\theta}_{\ga} {\theta}^{\ga}
+\bar{\eta}C*\bar k * \bar\varkappa* \bar{\theta}_{\pa}\bar{\theta}^{\pa}\right). \ee The second-order part
of $B_{2}$
\bee&&\nn
B_{2}\left(Z;Y;K\right)=-\Delta_{tw}^{*}\left(W_{1}*C-C* W_{1} \right) +C,\\
\nonumber
 &  & \Delta_{tw}C =0
\end{eqnarray}
contributes to  the equation for the $Z$-independent part $\go$ of $W_{2}$
\begin{equation}\label{W2}
\D_{ad}\omega=L(C)- {\Hh}_{ad}\left(W_{1}*W_{1}+i\eta
 B_{2}*  k *  \varkappa*  {\theta}_{\ga} {\theta}^{\ga}+i\bar{\eta}B_{2}*\bar k * \bar\varkappa* \bar{\theta}_{\pa}\bar{\theta}^{\pa}\right)
\end{equation}
with $L(C)$ \eq{LotC}. For the reader's convenience  manifest formulae for
$ {\Hh}_{tw}$, $\Delta_{tw}^{*}$, $ {\Hh}_{ad}$ and $\Delta_{ad}^{*}$ of \cite{Didenko:2015cwv}
are  collected in Appendix A.

\subsection{Current interactions}
\label{Currinter}
Schematically, Eqs.~\eqref{order2}   have the form
\be\label{int2}
 \D_{ad} \go + \go * \go-L(C) = G(w,\PPP)\,+Q(C,\go ),\qquad
\ee
\be
\label{int1}
\D_{tw}  C + [\go\,, C]_*=F(w, \PPP)   \q\quad
\ee
where $\!w=(\go_L,\bar\go_L,h)$ and the current
 $\PPP  $  is identified with the bilinear combinations of $C$
\be\label{PPP} \PPP (y^1{},y^2{};\bar y^1{},\bar y^2{};K|x): = C(y^1{},\bar y^1{};K|x)
 C(y^2{},\bar y^2{};K|x)\,.
 \ee
So defined $\PPP (y^1{},y^2{},\bar y^1{},\bar y^2{};K|x)$ verifies the {\it current
equation }
\be\label{tw2}
 \left( D_{L}   -{i} h^{\ga\pb}\Big (y^1{}_\ga \bar{y}^1{}_\pb- y^2{}_\ga \bar{y}^2{}_\pb
 -\p_1{}_\ga\bp_1{}_\pb+\p_2{}_\ga\bp_2{}_\pb\Big )\right)
 \PPP (y^1{},y^2{};\bar y^1{},\bar y^2{};K|x)=0
\ee
 at the convention that
 derivatives $\p_{1\ga}$($\bar \p_{1\dga}$) and $\p_{2\ga}$($\bar \p_{2\dga}$)
over the first and second  undotted(dotted)
spinorial arguments of $\PPP$ are defined to anticommute with
$k(\bar k)$:
\be\label{anticompj}
\p_2 A(w_1)k B(w_2) = - A(w_1)k\p_2 B(w_2)\,\q
\bp_2 A(w_1)\bar k B(w_2) = - A(w_1)\bar k\bp_2 B(w_2)\,.
\ee
(The presence of either $k$ or $\bar k$ in the first factor of $C$ leads to
the change of a relative sign between the sector of $Y_1$ and $Y_2$
in (\ref{tw2}) upon the Klein operator in the first factor of
$C(y^1{},\bar y^1{};K|x)$ is moved  through the second factor.)

By virtue of   Eq.~\eq{Csumkbark} the bilinear current
has the form
\be \label{Csumkbark'}
\PPP(Y^1;Y^2;K|x)=
\sum_{j,\,l=0,1 }
C^{j,1-j}(Y^1|x) k^j\bar k{}^{1-j}
  C^{l,1-l}(Y^2|x) k^l\bar k{}^{1-l}\,.
\ee
Nontrivial currents, that cannot be expressed via space-time derivatives of
the others, are identified with the primary components of the conformal module
realized by $\PPP(Y^1;Y^2;K|x)$ (see e.g. \cite{Gelfond:2015poa} and references
therein). These are the conserved currents  found originally in \cite{Berends:1985xx}

A simple but important fact, which follows
 from the analysis of    
 \cite{Gelfond:2015poa}, is that the $\go$-dependent
terms on the {\it l.h.s.} of Eq.~(\ref{int1}) and usual interactions with gauge invariant
currents contribute to different sectors of the equations. Namely, let $\pss$ be the spin of the field $C$ in the first term of
(\ref{int1}) while $s_1$ and $s_2$ be spins of the constituent  fields of $J\sim CC$.
Then the $\go$-dependent terms can be non-zero only at
$\pss<  s_1 +s_2\,.$

Hence  the $F$-terms in \eq{int1} corresponding to the usual gauge invariant
(\ie $\go$-independent)  current interactions are in the region
\be
\label{sss}
\pss\geq s_1 +s_2\,.
\ee
In other words, the gauge invariant currents $\PPP$ built from the zero-forms $C$
have spin $\pss$ obeying (\ref{sss}).
Note that this conclusion is in agreement with the results of \cite{Smirnov:2013kba}
where the currents with spins beyond the region (\ref{sss}) were built in terms
of the gauge connections $\go$. On the other hand, this argument does not apply
to the matter fields of spins $\pss<1$ having no associated gauge fields represented by
$\go$ on the left-hand-side of (\ref{int1})  in which case currents $\PPP$ (which are now
gauge invariant for a trivial reason) can be built
 from the zero-forms $C$ obeying $s_{1,2}\leq\half.$

Recall, that the equations in the one-form sector can be decomposed into the sum of
spin-$\pss$ eigenvectors  of the operator
\be\label{hats} \hat{s}=     y {}^\ga \p_\ga
 +\by {}^\pa \bp_\pa\,
 \ee with positive integer eigenvalues $2 (\pss+1)$.

To analyze local current deformations it is
  convenient  to use the mutually commutative  algebras  $\slv$   (vertical) {  and}
  $\slh$ {  (horizontal)}     \cite{Gelfond:2010pm},    which   are  dual to the
   rank-two covariant derivative (\ref{tw2}), mapping a solution of the current equation to a solution.   Thus,
  solutions of the  current equation   form a ${\slv}\otimes{\slh} $-module.
 In this paper  we will use  $\slv$ with the  generators
\bee \label{slv}\ls
f_+&=&y^1{}^\gga y^2{}_\gga - \bp_1{}_\pga \bp_2{}^\pga\,,\quad
f_-=\p_1{}_\gga \p_2{}^\gga-\by^1{}^\pga \by^2{}_\pga\,,\quad
 \\\nn
   f_0{} &=&y^1{}^\ga\p_1{}_\ga
+ y^2{}^\ga\p_2{}_\ga-\by^1{}^\pa\bp_1{}_\pa-\by^2{}^\pa\bp_2{}_\pa\q
\,\quad\\ \nn
[f_0,f_-]&=&-2f_-\q[f_0,f_+]=\,2f_+ \q [f_+,f_-]=  f_0{}.\eee
Note that the following  useful formula
\be\label{commfkf}
[f_+\,,\exp ( a f_-) \,]= a\exp (a f_-) (f_0-  a f_-  )
   \ee
  is a   consequences of Eq.~(\ref{slv}) and the  relation
 \be\nn
 \left[A,\exp B\right]=\int_0^1 dt \exp t B \left[A, B\right] \exp (1-t) B\,.
 \ee

 Any function (formal series) $F(y^1,y^2,\by^1,\by^2)$ can be decomposed into the sum of
eigenvectors $F_\hf(y^1,y^2,\by^1,\by^2)$ of the operator  $f_0$ (\ref{slv}) with   integer
eigenvalues $2\hf$,
\be\label{difJn}   F =\sum_\hf F_\hf\q
f_0F_\hf(y^1,y^2,\by^1,\by^2)=2\hf F_\hf(y^1,y^2,\by^1,\by^2).\ee
   A projection of $F$
to the eigenvector  with
eigenvalue   $2\bf \hf$ will be denoted  $F_\hf$.
 Eigenvectors $F_\hf(y^1,y^2,\by^1,\by^2)$ form an
$\slv$-module. From \eq{slv} it follows that
 \be\label{difJnfpm}
 \big[(f_-)^k{F}  \big] _{ \hf-k}=(f_-)^k F_\hf \q
  \big[(f_+)^k{F} \big]_{   \hf+k}=(f_+)^k F_\hf\,.
\qquad\,\,\,\ee
For a bilinear current  $\PPP_\hf,  $ $ \hf$
is the   sum of helicities  of the constituent fields
  \be\label{bilinhfCC}  \PPP_\hf(y^1,y^2,\by^1,\by^2;K|x):=
  \sum_{ {{h_1}+ {h_2}=\hf} }
  C_{h_1} (y^1,\by^1;K|x)\,C_{h_2} (y^2,\by^2;K|x)\, ,\ee
    where $C_{h_j}$  is a  helicity-$h_j$ field.
 Note that $\hf$ should not be confused with the
helicity of $\PPP_\hf$. For instance scalar constituent fields with $\hf=0$ generate currents
of any helicity (spin).

\subsection{Quadratic corrections in the zero-form sector}
\label{Quadraticc}
In this section we summarize results of \cite{Vasiliev:2016xui} on the computation of
current interactions in the zero-form sector.

 Quadratic deformation to equations on the zero-forms $C$ resulting
 from \eqref{eq:HS_1}, \eqref{eq:HS_2} by virtue of the procedure explained in
 Section \ref{Second order} has the form \cite{Vasiliev:2016xui}
\be
\label{HH}
 \D_{tw} C+ [\go\,, C]_*+  \Hhh_\eta (\PPP)+\Hhh_{\bar{\eta}} (\PPP)=0\,,
\ee where $\go$ stands for the first-order (\ie  not containing the
vacuum part $w$) part of  the $Z$-independent part of HS connection\be
\label{inda}
\go(Y;K|x)=W(0;Y;K|x)\,,\qquad C(Y;K|x)= B(0;Y;K|x)\,.
\ee
  The quadratic deformation is \bee
\label{CC}
\Hhh_\eta (\PPP) =&&\ls-\f{i}{2}  \eta\int  dS dT 
\exp i S_A T^A \int^1_0 d\tau \nn\\
&& [ h(s, \tau \bar y - (1-\tau) \bar t)
\PPP( \tau s,-(1-\tau)y +t, \bar y +\bar s,\bar y+\bar t;K)\nn\\
&&- h (t, \tau \bar y  - (1-\tau)\bar s) \PPP
((1-\tau)  y +  s, \tau  t, \bar y+ \bar s,\bar y+\bar t;K)]*k \,\q
\eee
\bee
\label{barCC}
\Hhh_{\bar\eta} (\PPP) =&&\ls-\f{i}{2} \bar \eta \int  dS dT
\exp i S_A T^A \int^1_0 d\tau
\nn\\
&&\ls\ls[  h (\tau  y  - (1-\tau) t,\bar s) \PPP
(y+s,y+t, \tau \bar s, -(1-\tau)\bar y +\bar t;K)\nn\\
&&\ls\ls- h ( \tau  y  - (1-\tau) s,\bar t) \PPP
(y+s,y+t,(1-\tau) \bar y + \bar s, \tau \bar t);K]*\bar k\,,\qquad
\eee
where
\be\label{hMV}h( a ,\bar b):=h^{\ga\pb}a_\ga\bar b_\pb\,.
\ee

 Formulae (\ref{CC}), (\ref{barCC}) follow from (\ref{C1}) and  (\ref{adj})-(\ref{antidz}).
The integration over $S$ and $T$ in (\ref{CC}), (\ref{barCC})  brings infinite tails of contracted indices
which, by virtue of the free unfolded equations (\ref{CON2k}), (\ref{tw}),  effectively  induce
an infinite expansion in higher space-time derivatives of the constituent
fields.
The field redefinition of \cite{Vasiliev:2016xui}
\be
\label{red}
C(Y;K|x) \to   C(Y;K|x)+ \Phi_{ \eta}(\PPP) +\Phi_{\bar \eta}(\PPP) \ee
with
\be\label{redef1} \Phi_{ \eta} (\PPP)
= \half\eta \int  dS dT  \exp i S_A T^A \int d^3\tau
\prod_{i=1}^3
\theta(\tau_i) \delta'  (\gs)
 \PPP( \tau_3 s+\tau_1y,t-\tau_2 y , \bar y +\bar s,\bar y+\bar t;K)*k\,,
\ee
\be
 \Phi_{\bar \eta} (\PPP)
=\half\bar\eta \int  dS dT  \exp i S_A T^A \int d^3\bgt
\prod_{i=1}^3
\theta(\bgt_i) \delta'(\bar \gs)
 \PPP(  s+ y,t+ y , \bgt_1\bar y +\bgt_3\bar s,-\bgt_2\bar y+\bar t;K)*\bar k\,,
\ee
\be
\gs= 1-\sum_{i=1}^3 \gt_i \q  \bar \gs = 1-\sum_{i=1}^3 \bgt_i\,,
\ee
replaces $\Hhh$ in equation (\ref{HH}) by $ \Hhh^{loc}$
 \be\label{hh}
 \Hhh^{loc}_{\eta \,cur} (\PPP)=
 \frac{1}{2}\eta
  \exp{(i [ \bp_1{}_\dgb  \bp_2{}^\dgb ])}
\int_0^1 d\tau h( y, (1-\tau) \bp_1-\tau\bp_2  )
\PPP(\tau y , -(1-\tau) y , \bar y   , \bar y ;K) *k\,,
\ee
\be\label{barhh}
 \Hhh^{loc}_{\bar \eta\,cur}  (\PPP)=
  \frac{1}{2}{\bar \eta}   \exp i [  \p_1{}_\gb   \p_2{}^\gb ]
\int_0^1 d\tau h((1-\tau)  \p_1- \tau   \p_2{}  ,\by)
\PPP(   y   ,  y ,\tau \bar y , -(1-\tau) \bar y;K ) *\bar k\,.
\ee
This current deformation is local since, containing only one type of contractions
between  spinor indices, for any
given spins $\pss,s_1,s_2$ it contains a finite number of derivatives.

Our goal is to extend these results to current deformation  in the one-form sector.

 \section{Main results}
\label{Main results}

The main result of this paper consists of the derivation of local current
interactions from the nonlinear HS equations. We actually obtained the two
forms of local current interactions, referred to as {\it natural} and {\it canonical}. The natural
form is simpler but contains some higher-derivative terms. The canonical form
is a bit more involved, but contains currents with the minimal number of derivatives.
Being related by a local field redefinition, the two forms are physically
equivalent. For the reader's convenience we present here both of them.
Let us stress that  both of these current deformations are proportional
to $\eta\bar\eta$, being independent of the phase of $\eta$.

The {\it natural form} of deformed equations is 
\be \label{W2hhloctor}
\D_{ad}\omega +\go*\go =
L(C ) + Q(C,\go )+  {\GGG}^{loc}_{\eta\bar\eta }(\PPP)\,,\,
\qquad\ee\be
 \label{W2hhloctortw}
 \D_{tw} C+ [\go\,, C]_*=
 - \Hhh^{loc}_{\eta\,cur} (\PPP)-\Hhh^{loc}_{\bar{\eta}\,cur} (\PPP) \,
   \ee
with $L(C )$ \eq{LotC},
   $  \Hhh^{loc} $   \eqref{hh},
  \eqref{barhh},   \bee\label{wc1111}
\ls  Q(C,\go )= &&  \!\! \ls\eta   \int dSdT\exp(iS_A T^A )
    \intop_{0}^{1} {d\gt}\nn
 \Big\{h(    t ,   \gt\bt-\bs  )   \go((1-\tau)y+s,\by+\bs )
 C( \gt t,\by +\bt ;K|x)
\\ &&+h(  s ,\bs \gt
  - \bt )
  C(\!-\!\gt s,\by +\bs ;K|x)
   \go(\!-\!(1-\tau)y\!-\!t ,\by\!+\!\bt )\Big\} \stt k +c.c.
 \eee
 and
\bee \label{DXloc}
   \GGG^{loc}_{\eta\bar\eta } \,\,=   \frac{i}{8}\eta\bar\eta
\int       \f{ d^4\gt}{  \gt_4^2 }    \gd(1-\gt_3-\gt_4)\gd'(1\!-\!\gt_1\!-\!\gt_2)
   \theta(\gt_1) \,\theta(\gt_2) \,    \theta( \gt_3)\,\,  \theta( \gt_4)
  \qquad\\ \nn
  \left\{\rule{0pt}{11pt}\right. \bar{H}^{\pa\pb}
\bp_\pa\bp_\pb%
  \exp i  \gt_3\bp_1{}_{\pa}\bp_2{}^{\pa}
    \PPP(    \gt_1y,  -  \gt_2 y ,   \gt_4 \gt_2\by,-\gt_4 \gt_1\by;K)
 \qquad
\qquad\\ \nn    +   H^{\ga\gb}
\p_\ga\p_\gb%
        \exp i  \gt_3\p_1{}_{\ga}\p_2{}^{\ga}
  \, \PPP(   \gt_4\gt_1y,  - \gt_4 \gt_2 y ,    \gt_2\by, -\gt_1\by;K) \left.\rule{0pt}{11pt}\right\},
\qquad\eee
where $\p_{1\ga}, \p_{2\ga},\bar\p_{1\dga}$ and $\bar\p_{2\dga}$ are, respectively, derivatives over the first,
second, third and fourth spinorial arguments of $J$ with upper indices. Note that the
$\go*\go$ and $\go C$ terms
(\ref{wc1111}) are
local since  $\go(y,\bar y)$ is  polynomial in $y$ and $\bar y$ for any finite spin.

Being defined for
 currents   respecting condition \eq{sss}, the  canonical form of deformed equations is
\bee \label{W2hhloc0}
\D_{ad}\omega+ \go * \go &=&
L(C)+ Q(C,\go )+{\GGG}^{loc}_{\eta\bar\eta }\Big|_{\pss<s_1+s_2}(\PPP)\,+ {\GGG}^{can}_{\eta\bar\eta }(\PPP),\,\qquad
\\
 \label{W2hhloc0tw}
 \D_{tw} C+ [\go\,, C]_*&=&   - \Hhh^{loc}_{\eta\,cur} (\PPP)-\Hhh^{loc}_{\bar{\eta}\,cur} (\PPP) \,
 +\D_{tw}   B^{sum} (\PPP) \,.
  \eee
  The current deformation in the zero-form sector consists of two parts. The first one,
   $ -( \Hhh^{loc}_{\eta\,cur} (\PPP)+\Hhh^{loc}_{\bar{\eta}\,cur} (\PPP))$,  is the same as in
      Eq.~\eqref{W2hhloctortw}.  The second one defined in Eq.~\eq{sdvigSint}
does not contribute to the dynamical equations considered in Section \ref{contribution}
  except for  the spin-one case  (see Section \ref{s=1}).

 The current deformation in the  one-form sector consists of two parts. The first one,
$ {\GGG}^{loc}_{\eta\bar\eta }\Big|_{\pss<s_1+s_2}(\PPP)\, $,
is the projection of  $ {\GGG}^{loc}_{\eta\bar\eta }$
      \eqref{DXloc} to the region of gauge dependent deformation. The second one,
      ${\GGG_{\eta\bar\eta }^{can} }(\PPP)$, which  is explicitly defined in Eq.~\eq{canonical ressub==}, is just the gauge invariant
      deformation of \cite{Gelfond:2010pm}.

The  \emph{canonical current deformation} in the one-form sector
  ${\GGG}^{can}_{\eta\bar\eta }(\PPP)$ 
  is
\bee
\label{canonical ressub==}
 \ls{\GGG_{\eta\bar\eta }^{can} }(\PPP)\!\!&=&\!\!\f{i}{8}\eta\bar\eta
\int \f{d^2\gr}{\gr_1 }\f{ d^4\gt}{ \gt_4^{2}}
     \gd'(1-\gr_1-\gr_2)
   \gd (1-\gt_3-\gt_4)\gd'(1\!-\!\gt_1\!-\!\gt_2) \qquad \\ \nn &&\Upsilon \left(1\!-\!\f{\gd(\gr_2)}{2}\right)
 \oint \f{dw}{2\pi i w }
      w  ^{- 2|\pss|_2} \int  {d\bs d\bt} \exp  \big (  i   \bs_\pga \bt{}^\pga)
  \left[\rule{0pt}{12pt}\right.
 \bar H^{\pa\pb}   {\bp_\pa \bp_\pb}
 \qquad
 \\ \nn
 &&
 \PPP(     \gt_1y    w^{\!-\!1} ,
  \!-\!   \gt_2 y    w^{\!-\!1}  ,
  (\gr_1\gt_4\gt_2\by  \!+\!(\gr_1\gt_3 +
  \gr_2  w^{\!-\!2})  \bs) w   ,(
\!-\! \gr_1\gt_4 \gt_1\by \!+\!  \bt) w  ;K)
\rule{25pt}{0pt}
\\ \nn &&   
+
\int  ds dt H^{\ga\gb} {\p_\ga \p_\gb}(\gr_1)^{ 2|\pss|_2}
   \exp i\big(   s_\gga  t^\gga(   w^{\!-\!2} \gr_2 \gr_1 \gt_3-1)
   + \gt_4(\gr_1  \gt_1t_\gga   +\gt_3^{-1} \gt_2 s_\gga ) y^\gga
    \big)
\\ \nn && 
\PPP(     s  w^{\!-\!1} ,
     \gr_1 \gt_3  t    w^{\!-\!1}  ,
   (\gt_2\by   \!+\!  \bs)w   ,
(\!-\!   \gt_1\by \!+\! w^{\!-\!2} \gr_2 \bt) w   ;K) \left.\rule{0pt}{12pt}\right] +c.c.\,,
\eee     where $\p_\ga:=\f{\p}{\p y^\ga}$, $\p_\dga:=\f{\p}{\p \bar y^\dga}$, and the measure $\dis{\f{dw}{ 2\pi i w}
    w^ {-|2\pss|_{ 2}} }$
    differs for
bosonic currents with integer $\pss$ ($|2\pss|_{ 2}=0$) and
fermionic ones with half-integer $\pss$ ($|2\pss|_{ 2}= 1$).\rule{0 pt}{15pt}
Note  that the factors of $(1-  w^{\!-\!2} \gr_2 \gr_1 \gt_3)^{-1}$ resulting from
the Gaussian integration over $s_\ga$ and $t_\ga$ should be expanded in power series
in $w^{\!-\!2} \gr_2 \gr_1 \gt_3$.

As discussed  in Section \ref{examps=s0j}, formula (\ref{canonical ressub==})
reproduces the two types of $4d$ cubic vertices found by Metsaev in \cite{Metsaev:2005ar}.
The vertices
containing in $J$ the constituent fields  of  helicities of the same sign
describe the $AdS$ deformation of the Minkowski Lagrangian vertices with $\pss+s_1+s_2$
space-time derivatives. Those with  constituent fields of opposite helicity signs
 describe the $AdS$ deformation of the Minkowski vertices with
$\pss+|s_1-s_2|$ derivatives (recall that we assume that $\pss\geq s_1+s_2$).
Moreover, as  shown in \cite{misuna}, the resulting cubic  vertices precisely
reproduce the coefficients found by Metsaev in \cite{Metsaev:1991mt} from the analysis
of quartic vertices.

Note that our vertex contains both parity even and
parity odd parts, which appear in HS models with general $\eta$ upon transition to the genuine Weyl tensors as explained in \cite{Didenko:2017lsn}. More precisely,
this is true for the vertices with $\pss+s_1+s_2$ derivatives while those
with the minimal number of derivatives remain parity even for any $\eta$. Note that parity-odd vertices in four dimensions were considered in \cite{Boulanger:2005br} for  spin three
and in \cite{Conde:2016izb} for general spin.

\section{Derivation details }
\label{nonlincorone}

 \subsection{Summary of main steps}
\label{Scheme}

Before going into details of derivation of our results in the rest of this section,
we briefly summarize the main steps.

In Section \ref{nonlinetaeta} it is shown that the $C^2$-deformation
in the  sector of  one-forms resulting from the standard approach to nonlinear HS equations  is
 ${\HhhH}= {\HhhH}_{\eta^2}+ {\HhhH}_{\bar\eta^2}$ \ie
 ${\HhhH}_{\bar\eta\eta}=0$.
The bilinear field redefinition in the zero-form sector (\ref{red})
induces via the linear part    $L(C )$ \eq{LotC} of the Central-on-shell theorem the quadratic correction
 $ \GGG :=\GGG_{\eta^2  }+\GGG_{\eta\bar\eta }+\GGG_{\bar\eta^2 }$
so that Eq.~(\ref{W2}) acquires the form
\bee\label{W2hhbezexact}
\D_{ad}\omega+ \go * \go =L(C) - {\HhhH}_{ \eta^2}- {\HhhH}_{\bar\eta^2}+\GGG_{\eta^2  }
+\GGG_{\eta\bar\eta }+\GGG_{\bar\eta^2 }+  Q(\go\,, C)\,.
\eee
Here the nonzero $\go C$-deformation resulting from the nonlinear HS equations
in the  sector of  one-forms can be represented in the form
\bee\nn
Q(C,\go) = - {\Hh}_{ad}\Big[  -i\omega *\Delta_{ad}^{*}
\Big(C* \eta\gamma \Big)
-i\Delta_{ad}^{*}\Big(
C* \eta\gamma\Big) *\omega   \Big] +c.c.
\q\eee
 where
\be
\gga=  k *  \varkappa*  {\theta}_{\ga} {\theta}^{\ga}\q
\bar\gga=\bar k * \overline\varkappa* \bar{\theta}_{\pa}\bar{\theta}^{\pa} \,.
\ee
The computation of $Q$ is straightforward leading to \eq{wc1111}.

Then in Section \ref{nonlinetaeta}  it is shown that
\be\label{zeroetaeta1}
- {\HhhH}_{ \eta^2}- {\HhhH}_{\bar\eta^2}+\GGG_{\eta^2  } +\GGG_{\bar\eta^2 }
=\D_{ad} (\widetilde{\Omega}+\Psi)
\ee
with   forms  $ \widetilde{\Omega}$ \eq{gogo2til} and $ \Psi $ \eq{Psi}.
Upon the field redefinition $\go:=\go-(\widetilde{\Omega}+\Psi) (\PPP)$, the remaining
quadratic terms in the one-form sector turn out to be
 proportional to $\eta\bar \eta$
 \be\label{etabetashift'}
 \GGG_{\eta\bar\eta }(\PPP)=\frac{i}{8}\eta\bar\eta
 \int  {dS dT}  \exp i S_A T^A \int d^3\bgt    d^3\gt
\ee\be \nn \left\{\,\Upsilon\,
\gd(\gs)\gd'(\bar \gs)\gd(\gt_1)\gd(\gt_2)
\bar{H}^{\pa \pb}\bp_\pa\bp_\pb
    + \,\Upsilon\, 
  \gd'(\gs)\gd(\bar \gs)\gd(\bgt_1)\gd(\bgt_2)
H^{\ga \gb}\p_\ga\p_\gb
   \right\}\ee\be \nn
   \PPP(  \gt_3 s+\gt_1y;t -\gt_2 y ,     \bgt_3\bs+ \bgt_1\by  ; \bt-\bgt_2 \by ;K)
    \,,\qquad\ee
     where, abusing terminology, here and below we use the shorthand notation   $\Upsilon$ for a product of $\theta(\gt )$
for all necessary homotopy parameters $\gt $, $\bgt $, $\gr $ \etc, \ie those
to which no $\gd^{(m)}(\gt)$ \,, $\gd^{(k)}(\bgt)$\,, $\gd^{(n)}(\gr)$
\ldots is associated,\\
\bee  \label{Thetas}
  \Upsilon:= \prod _j
  \theta(\gt_{k_j})  \theta(\bar{\gt}_{i_j})  \theta(\gr_{l_j})\dots  .\qquad
\eee
In Section \ref{etabaretaloc} such  $X(\PPP)$ \eq{sourceredifeta0} is found  that $
\D_{ad}X = \GGG_{\eta\bar\eta } - \GGG^{loc}_{\eta\bar\eta }
$
with $\GGG^{loc}_{\eta\bar\eta }$ \eq{DXloc}. Upon the field redefinition
$\go\to\go- X(\PPP)$, the  quadratic terms
in the one-form sector aquire the local form
$\GGG^{loc}_{\eta\bar\eta }$. However, as explained in Section \ref{decomp2}, it
 contains higher-derivative terms with the
coefficients divergent in the flat limit. Finally, in
 Section \ref{Canonical deformations2} we find such  %
 local one-form   $\Lambda^{sum}$
and zero-form $B^{sum}$      that,
upon the field redefinition
\be\label{redifcan}\go\to\go+ \Lambda^{sum}(\PPP) \q C\to C+ B^{sum}(\PPP) \,,\ee
the quadratic terms in the one-form sector take gauge invariant
  {\it canonical current deformation} form
$ {\GGG}^{can}_{\eta\bar\eta }$ with the minimal number of derivatives,
that admits  a proper flat limit.

\subsection{Field redefinition in the $\eta^2$ sector}
\label{nonlinetaeta}

According to (\ref{W1}), (\ref{W2}),
the  $C^2$-deformation resulting from the nonlinear HS equations
in the  sector of  one-forms can be represented in the form
\bee \label{Hhh}
 {\HhhH}\ls\,\, &:=&
 {\Hh}_{ad}\left\{\Delta_{ad}^{*}\left(C*
\left(\eta\gamma+ \bar{\eta} \bar{\gamma} \right)\right)*\Delta_{ad}^{*}
\left(C*\left(\eta\gamma+ \bar{\eta} \bar{\gamma} \right)\right)\right\}\\ \nn  &+&{\Hh}_{ad}\left\{
 \Delta_{tw}^{*}\left[\Delta_{ad}^{*}\left(
C*\left(\eta\gamma+ \bar{\eta} \bar{\gamma} \right)\right)\,,C   \right]_*
 *\left(\eta\gamma+ \bar{\eta} \bar{\gamma} \right)\right\}\,.
\eee

Decomposing $ {\HhhH}$ (\ref{Hhh}) in powers of $\eta$ and  $\bar{\eta}$ as ${\HhhH}= {\HhhH}_{\eta^2}+{\HhhH}_{\eta\bar\eta}+ {\HhhH}_{\bar\eta^2}$
it is not hard to see   by virtue of (\ref{adj})-(\ref{antidz}) that ${\HhhH}_{\eta\bar\eta}=0$.

Consider
 \be\label{summaetaeta}
{\HhhH}_{ \eta^2}= \eta^2\Hh_{ad} \left\{
\Delta_{tw}^{\sstt} \big(\big[\Delta_{ad}^{\sstt} \big( \CCC \stt
 \gga  { }  \big)\,, \CCC \big]_\stt\,\big) \stt \gga {} { }\!+\!
  \Delta_{ad}^{\sstt} \big( C \stt
 \gga  { }  \big)\,\stt  \Delta_{ad}^{\sstt} \big( C \stt  \gga \big)\right\},  \ee
  where by virtue of     (\ref{adj})-(\ref{antidz})
and (\ref{Klein}), (\ref{EQdef})\,
\bee\label{first1}
&&\Hh_{ad}  \left(
\Delta_{tw}^{\sstt} \big(\big[\Delta_{ad}^{\sstt} \big( \CCC \stt
 \gga  { }  \big)\,, \CCC \big]_\stt\,\big) \stt \gga {} { }\right)=
\!\f{i   }{8}
  \!\int\limits_0^1\! { d\gt }
 \!\!\int\!\!  d S\, d T\,   \exp(i S_A T^A)
\qquad \\
&& \nn
\Big[\exp(i   (\gt s  \!-\!t )_\ga y^\ga  )
\Big\{ \go_L{}^\ga{}_\gga \go_L{}^\gb{}^\gga  (\gt s  \!-\!t )_\ga(\gt s  \!-\!t)_\gb
\!+\! 2h_\gga{}^\pa  \go_L{}^\gga{}^{ \gb}\,(\bt\!-\!\bs)_\pa
(\gt s  \!-\!t )_\gb\Big\}\,  \qquad
 \\  \nn&&
\!-\!\big\{\exp(i   (\gt s \!-\!t )_\ga y ^\ga  )\!-\!1\big\}\bar{H}{}^{\pa\pb}(\bt\!-\!\bs)_\pa
  (\bt\!-\!\bs)_\pb\,
\Big]   \PPP ( \gt  s ,\,   t,\,  \by   \!+\!\bs,\, \by   \!+\!\bt;K|x)\q
   \eee
 \bee
 \label{HDCstDC}
\! &&\Hh_{ad}\left\{\Delta_{ad}^{\sstt} \big( C \stt
 \gga  { }  \big) \stt  \Delta_{ad}^{\sstt} \big( C \stt  \gga \big)\right\}  =
  \f{   1}{8} \!\int\! \! d S  d T         
  \!\!  \int\limits_0^1\! \!  d\gt_1
\!\! \int\limits_0^1\! d\gt_2 \exp i \Big (S_A T^A +(\gt_1{} s_\gb   \!-\!  \gt_2{} t_\gb) y^\gb\Big )\nn
 \\ &&
  \Big\{     \go_L{}^\ga{}_\gga \go_L{} ^{\gga\gb}  s_\gd t^\gd
\gt_1\gt_2      \big(2 t _\ga s_\gb
   \!-\!\gt_1 s_\ga s_\gb\!-\!\gt_2t_\ga t_\gb
 \big )
  \!-\!2 \go_L{}_{\ga}{}^{ \gb}h^{\ga\pb} {   \gt_1{}\gt_2 s_\gd t^\gd
(   t_\gb \bs_\pb\!+\! s_\gb \bt_\pb) }\\ &&
 \!+\!   2{  \go_L^{\ga \gga}h^{\gd \pb}  (\gt_1{}\gt_2{}\!-\!1)( \!
   \gt_2 s_\ga t_\gga  t_\gd\bt_\pb -\!\gt_1{}    t_\ga     s_\gga  s_\gd   \bs_\pb
)
 }
 \!-\!  {H}{}^{\ga\,\gb} s_\ga t_\gb (\bs_\pa\bt^\pa\!-\!2i) ( \gt_1{}\gt_2{}\!-\!1 )\nn\\&&
  \!-\!\overline{{H}}{}^{\pa\,\pb}  s_\ga t^\ga
 \big(( \gt_1{}\gt_2{}\!+\!1 )\bs_\pa\bt_\pb
 \!-\!\gt_1{}\bs_\pa\bs_\pb \!-\!\gt_2{}\bt_\pb \bt_\pa\big)\Big\}
\PPP( \gt_1{}s, \gt_2{} t ,\by\!+\! \bs ,\by \!+\!\bt;K|x)\q
\eee
 with $\PPP$ (\ref{PPP}).

First of all, the part of ${\HhhH}_{\eta^2}$ containing $\go_L^2 $  has to be eliminated.
To this end we set
\be\label{gogo2til}\widetilde{\Omega}   =   \go_L{}^{\ga\gb}
\Omega_{\ga\gb}\,,\ee
 \be \label{gogo2}
\Omega_{\gd\gb}=
  i
 \f{ \eta^2}{4}\!\!\int\! d S d T    s_\gd t_\gb      
    \int\limits_0^1 { d\gt_1{} }\gt_1{}\!\!
\! \int\limits_0^1 { d\gt_2{} }\gt_2
 \exp i ( S_A T^A +(\gt_1{} s_\gga -\gt_2{} t_\gga  ) y ^\gga)
 \PPP( \gt_1{}s,\gt_2{}  t,\by +\bs {},\by +\bt;K|x)\,
.\ee From   (\ref{znakH}), (\ref{Dad})
it follows that
  \be\label{DOmega}
  \D_{ad}\widetilde{\Omega}= \go_L{}^\gb{}_\gga \go_L{} ^{\gga\,\gd} \Omega_{\gb\gd}  +H^{\ga\gb}\Omega_{\ga\gb}
- \go_L{}^{\ga\gb}   \D_{ad}  {\Omega}_{\ga\gb}.\ee
Using the useful identity
\be \label{inttauF}
 \int\! ds dt
   \int\limits_0^1 { d\gt {} }\exp(i s_\ga t^\ga  )\Big\{
F(\gt  s)  \, (\gt )^{n-1}\big(n-2 -i s_\gb  t^\gb \big)-\f{\p}{\p \gt}
\big((\gt )^n F(\gt  s)\big)\Big\}=0\,,
 \ee
      by virtue of  (\ref{Dad}) and
 (\ref{tw})    it is not hard to see that  $\HhhH_{\eta^2}$   can be represented in the form
\bee\label{restrest}
 \HhhH_{\eta^2}=\D_{ad}\widetilde{\Omega}
  +  \D_{ad} \Psi +  {\HhhH'}_{\eta^2}\q
 \eee where\be \label{restrestetaeta}
  {\HhhH'}_{\eta^2}\!=\!  -i\f{  \eta^2}{8}\bar{{H}}{}^{\pa \pb}
 \! \int\limits_0^1\! { d\gt }\!\!
\int   \!\! d S  d T \exp (i   S_AT^A)
 (\bt-\bs)_\pa(\bt-\bs)_\pb
    \PPP(\gt  s ,    t,   \by   +\bs,  \by   +\bt;K)\,,
    \ee
    \be\label{Psi}
\Psi=-i\f{  \eta^2}{4}    h(s,\bs)
 \int_0^1\!d\gt_1\gt_1{}\! \int_0^1\!d\gt_2\!\int\! dS dT\exp
  i (  S_AT^A +(\gt_1{} s -\gt_2{}  t   )_\gga y ^\gga )
\PPP(\gt_1{}s,    \gt_2{}t,\by +\bs {},\by +\bt ;K). \qquad\,
 \ee

 It should be stressed that both in $ {\HhhH}_{\eta^2}$ (\ref{summaetaeta}) and
in $\Psi$ (\ref{Psi}) the dependence on the right spinors $\bar y^\dga$ remains unaffected
by the homotopy integrals, \ie these variables are only affected by the star product
in the right sector. This Ansatz is specific for the separation of
variables approach applied in \cite{Vasiliev:2016xui} to the zero-form sector, where
it was shown to lead to the unique local solution, and extended in this paper to
the one-form sector.  Modulo gauge transformations and local changes of variables,
field redefinition (\ref{Psi}) is also the only one  respecting the separation of variables
and giving the local result.

Analogously, modulo $\D_{ad}$-exact terms, the $\bar\eta^2$-deformation is
\be \label{restrestbetabeta}
  {\HhhH'}_{\bar\eta^2} =  -i\f{  \bar\eta^2}{8}{{H}}{}^{\ga \gb}
 \! \int\limits_0^1 { d\bgt }
  \int    d S  d T
\exp (iS_AT^A)
  (t- s)_\ga( t-s)_\gb   \PPP( y+ s , y+   t,   \bgt\bs,   \bt;K|x)\,.\qquad
\ee
Thus, upon the field redefinition $$\go(y,\by;K|x)\to  \go(y,\by;K|x)+\widetilde{\Omega}+\Psi$$
 Eq.~(\ref{W2}) acquires the form
\begin{equation}\label{W2hh}
\D_{ad}\omega + \go * \go=L(C )+Q(C,\go )- {\HhhH'}_{ \eta^2}- {\HhhH'}_{\bar\eta^2}\,.\quad
\end{equation}

 The field redefinition  in the zero-form sector (\ref{red}) 
induces the quadratic correction  on the  \rhs of  (\ref{W2hh}) by virtue of \eq{LotC}
\be\label{Gamma} \GGG:=\f{i}{8}\bar\eta  H^{\ga\gb} \p_{\ga}\p_{\gb}
\Big\{\Phi_{ \eta} +\bar{\Phi}_{\bar \eta}\Big\}(\PPP)(y,0;K| x) +
\f{i}{8}\eta
 \bar{H}^{\pa\pb} \bp_{\pa}\bp_{\pb}
 \Big\{\Phi_{ \eta}+\bar{\Phi}_{\bar \eta}\Big\}(\PPP)(0,\overline{y};K| x)\q
\ee
which can be decomposed as  $ \GGG :=\GGG_{\eta^2  }+\GGG_{\eta\bar\eta }+\GGG_{\bar\eta^2 }$.
Upon  integration over $\gt_1$ and $\gt_2$
\be \nn
 \GGG_{\eta^2}=\frac{ i}{8}\eta^2
 \int  {dS dT}  \exp i S_A T^A \int d \gt_3
\theta(\gt_3)\theta(1-\gt_3)
   \bar{H}{}^{\pa \pb}\bp_\pa\bp_\pb
     \PPP(  \gt_3 s ,t   ;      \bs+  \by  , \bt+  \by ;K|x)\,
 \ee
 just compensates  $- {\HhhH'}_{ \eta^2}$ \eqref{restrestetaeta} in (\ref{W2hh}).
      Hence,  the full $\eta^2$-deformation is zero.
      Analogously, the full $\bar\eta^2$-deformation is also zero.
 Hence the nonlinear deformation in the one-form sector takes the form
\begin{equation}\label{W2hh'}
\D_{ad}\omega + \go * \go=L(C )+Q(C,\go ) + {\GGG}_{ \eta \bar\eta }(\PPP),
\end{equation}
with $L(C)$ \eq{LotC}, $Q(C,\go )$ \eq{wc1111} and $\GGG_{\eta\bar\eta }(\PPP)$ \eq{etabetashift'}.
   Clearly, $\GGG_{\eta\bar\eta }(\PPP)$ is not local.

The fact that there exists a field redefinition of the one- and zero-forms bringing
 the $\eta^2$-terms in the one-form sector  to zero  is not {\it a priori}
 obvious. Remarkably to reach this result one has just to apply
 the  shift in the zero-form sector  found in  \cite{Vasiliev:2016xui}.
In other words, the alternative way to deduce the shift of \cite{Vasiliev:2016xui} is
to demand that the  $\eta^2$-terms in the one-form sector should be zero.

 \subsection{From nonlocal to local deformation in the $\eta\bar\eta$ sector }
\label{etabaretaloc}
   Now we show, that there exists a proper field
 redefinition bringing   deformation (\ref{etabetashift'}) to the local form.
Due to the gauge freedom for one-form HS gauge fields there exist many equivalent
representatives for the same local current.
An alternative field redefinition that contains integration over a simplex
in the space of homotopy parameters is presented in Appendix B.
It should be stressed that the field redefinition found in this section is unique modulo
gauge transformations and further local transformations because the zero-form equations
found in \cite{Vasiliev:2016xui} are demanded to be unaffected.

The problem will be solved in two steps. First, in this section we will
introduce a $\D_{ad}$-exact shift bringing the deformation to the
local form. Second, in Section \ref{Flat limit }, a local field redefinition eliminating both the terms divergent
in the flat limit and  those contributing to the  torsion-like HS curvature
will be found  in the sector of gauge invariant currents obeying (\ref{sss}).

   Let
\bee\label{sourceredifeta0}
  X (\PPP)= \frac{ i}{8}\eta\bar\eta\,
  \int    d^3\gt d^3 \bar\tau \Upsilon \gd(1\!-\!\gt_3\!-\!\gt_2)
  \gd(1\!-\!\bgt_3\!-\!\bgt_2)\gd'(1\!-\!\gt_1\!-\!\bar\tau_1)
  \,h(\p,\bp)\qquad\\ \nn
  \f{(1\!-\!\gt_3\bgt_3)}{ \gt_2 \bgt_2 }
  \exp i\big(  \gt_3\p_1{}_{\ga}\p_2{}^{\ga}+\bgt_3\bp_1{}_{\pa}\bp_2{}^{\pa}
  \big) \PPP(   \gt_2 \gt_1y,  \!-\!\gt_2 \bar\tau_1 y ,   \bgt_2 \bar\tau_1\by,\!-\!\bgt_2 \gt_1\by;K|x) \,  \qquad
\eee
with $\Upsilon$ \eq{Thetas}.
Using (\ref{znakH}), straightforward differentiation  yields
 \bee\label{DadX}
 \D_{ad}X =  \frac{ i}{8}\eta\bar\eta\,
  \bar H^{\pa\pb}
\bp_\pa\bp_\pb%
   \int_0^1 d \gt_3  \f{ {\p}}{\p \gt_3}\int d  \bgt^3
   d\gt_1 \Upsilon   \gd(1\!-\!\bgt_3\!-\!\bgt_2)\gd'(1\!-\!\gt_1\!-\!\bar\tau_1)
    \Big\{ \f{(1\!-\!\gt_3\bgt_3)^2 }{ \bgt_2^2 }
 \qquad \\ \nn
  \exp i\big(  \gt_3\p_1{}_{\ga}\p_2{}^{\ga}+\bgt_3\bp_1{}_{\pa}\bp_2{}^{\pa}
  \big) 
  \PPP(   (1\!-\!\gt_3)\gt_1y,  \!-\!(1\!-\!\gt_3) \bar\tau_1 y ,
  \bgt_2 \bar\tau_1\by,\!-\!\bgt_2 \gt_1\by;K|x)\Big\}
   \nn    + c.c.
  \eee
 Note that the $\bgt_2$-poles
 in Eq.~(\ref{DadX})  are fictitious  due to the differentiations
 $\bp_\pa\bp_\pb$.

 Hence performing integration over $\gt_3$ one obtains
 \be
\D_{ad}X(\PPP) = \GGS^{nloc}(\PPP) - \GGG^{loc}_{\eta\bar\eta }(\PPP)
\ee
with
\bee\label{GGSN}  \GGS^{nloc}(\PPP)=\frac{i}{8}\eta\bar\eta
\int      d^4\gt \Upsilon    \gd(1-\gt_3-\gt_4)\gd'(1\!-\!\gt_1\!-\!\gt_2)
   \qquad\qquad\\ \nn \Big\{\bar H^{\pa\pb} \bp_\pa\bp_\pb%
   \exp i   \big( \p_1{}_{\ga}\p_2{}^{\ga}+\gt_3\bp_1{}_{\pa}\bp_2{}^{\pa}
    \big)\PPP(   0, 0,   \gt_4 \gt_2\by,-\gt_4 \gt_1\by;K|x)
 \qquad
 \\ \nn  +   H^{\ga\gb}
\p_\ga\p_\gb%
   \exp i\big(  \gt_3\p_1{}_{\ga}\p_2{}^{\ga}+ \bp_1{}_{\pa}\bp_2{}^{\pa}
  \big) \PPP (   \gt_4\gt_1y,  - \gt_4 \gt_2 y ,  0,0;K|x)
\Big\} \, \eee
and  $ \GGG^{loc}_{\eta\bar\eta } (\PPP)$  \eq{DXloc}.
 By virtue of    the following simple formula\bee\label{int3int4}
\int d^3\gt \gd'(1-\gt_1-\gt_2-\gt_3)\theta(\gt_1) \,\theta(\gt_2) \,\theta(\gt_3)\,
f(\gt_1 \,, \gt_2\,,\gt_3) \qquad
\\ \nn=\int d^4\gt \gd'(1-\gt_1-\gt_2 )
\gd (1-\gt_3-\gt_4)\theta(\gt_1) \,\theta(\gt_2) \,\theta(\gt_3)\, \theta(\gt_4)\,
f(\gt_1\gt_4 \,, \gt_2\gt_4\,,\gt_3)
\qquad\eee
 $\GGS^{nloc}$ \eqref{GGSN}
coincides with $\GGG_{\eta\bar\eta }$ (\ref{etabetashift'}). Therefore, by a field redefinition
\be\label{sdvigomega1}
\go\to\go- X(\PPP)
\ee
with $ X(\PPP)$     (\ref{sourceredifeta0}),
 the  $\eta\bar\eta$-current deformation
 in the one-form sector is reduced to  $\GGG^{loc}_{\eta\bar\eta }(\PPP) $.

The following comment is now in order. The deformation $\GGG_{\eta\bar\eta }$ (\ref{etabetashift'})
consists of two terms. One can check that each of these terms is $\mathcal{D}_{ad}$-closed.
Originally, we anticipated that each of these terms can be brought to the local form
by a field redefinition  in the form of some homotopy integral.
However, we failed to proceed this way. This  is in agreement with the final
field redefinition represented by a single term (\ref{sourceredifeta0}), providing one more
evidence of the uniqueness of  the proposed scheme.

Being local,  deformation (\ref{DXloc}) does not reproduce the {\it canonical}
deformation  of \cite{Gelfond:2010pm}.
Hence the difference between the two forms of local currents should be an improvement, \ie
 to bring  deformation (\ref{DXloc}) to the canonical form we have to perform a further local
 field redefinition.

\subsection{Derivation of the canonical  form of current interactions}
\label{Flat limit }
 \subsubsection{Flat limit rescalings}
 \label{flatlim}
To take the flat limit it is necessary to perform certain rescalings.
 To this end, following to \cite{33}, it is useful to introduce notations
$A_\pm$ and $A_0$ so that the eigenvalues of the helicity operator
$
\half \left (y^\ga\f{\p}{\p y^\ga} - \overline{y}^\pa
\f{\p}{\p \overline{y}^\pa}\right )
$
are positive on  $A_+(y,\overline{y}\mid x) $,
negative on  $A_-(y,\overline{y}\mid x) $, and zero on
$A_0(y,\overline{y}\mid x)$.
Using the decomposition
\be
\label{dec}
A(y,\overline{y}\mid x) =
A_+(y,\overline{y}\mid x) + A_-(y,\overline{y}\mid x)
+A_0(y,\overline{y}\mid x)\,,
\ee
the rescalings are introduced differently in the adjoint and twisted
adjoint modules
\bee
\label{resc}
\tilde{A}^{ad}(y,\overline{y}\mid x)=&&\ls
A_+(\lambda^{\half} y,\lambda^{-\half}\overline{y}\mid x) +
 A_-(\lambda^{-\half}y,\lambda^{\half} \overline{y}\mid x)
+A_0(y,\overline{y}\mid x)\,,\\ \nn
\tilde{\tilde{A}}^{tw}(y,\overline{y}\mid x)=&&\ls
A(\lambda^{\half}y,\lambda^{\half}\overline{y}\mid x)\,.
 \eee
For the rescaled variables, the flat limit $\lambda \to 0$ of
the adjoint and twisted adjoint covariant derivatives (\ref{Dad})
and (\ref{tw}) gives
\be
\label{adfl}
D_{fl}^{ad}\tilde{A}(y,\bar{y} \mid x)
= D^L \tilde{A} (y,\bar{y} \mid x) +
 \Big (h( y,\bp)
\tilde{A}_-(y,\bar{y} \mid x)
+h ( \p, \by) \tilde{A}_+(y,\bar{y} \mid x)
\Big ) \,,
\ee
\be
\label{fltw}
D_{fl}^{tw} \tilde{\tilde{A}}(y,\bar{y} \mid x) =
D^L \tilde{\tilde{A}}(y,\bar{y} \mid x) +i h( \p ,\bp)
\tilde{\tilde{A}}(y,\bar{y} \mid x)\,.
\ee
The flat limit of the unfolded massless equations results from
(\ref{CON1k})   via the substitution of
$D^L$ and $h^{\ga\pa}$ of Minkowski space along with the
replacement of $D^{ad}$ and $D^{tw}$
by $D^{ad}_{fl}$ and $D_{fl}^{tw}$, respectively.
The resulting field equations
describe free HS fields in Minkowski space. Let us stress that,  looking somewhat unnatural
in the two-component spinor notation, prescription (\ref{resc})
is designed just to give rise to the theory of
Fronsdal \cite{Frhs} and Fang and Fronsdal
 \cite{Frfhs} (for more detail see \cite{33}).

Note that, although  the contraction $\lambda\to 0$
with the rescaling (\ref{resc}) is consistent with the free
HS field equations,  negative powers
of $\lambda$ survive in the full nonlinear equations
 upon  rescaling (\ref{resc}), making
the Minkowski background  unreachable in the
nonlinear HS gauge theories of \cite{Fradkin:1987ks,more,Vasiliev:2003ev}. Nevertheless, the
HS interactions with gauge invariant currents considered in this paper
do admit a proper flat limit.

\subsubsection{Current decomposition}
\label{decomp2}

Firstly let us represent $\GGG^{loc}_{\eta\bar\eta }$ \eqref{DXloc} as
\be\label{tworegloc}
\GGG^{loc}_{\eta\bar\eta }(\PPP)=\GGG^{\ge\,loc}_{\eta\bar\eta }(\PPP)
+\GGG^{<\,loc}_{\eta\bar\eta }(\PPP)\,,
\ee
where
\be\nn\GGG^{\ge\,loc}_{\eta\bar\eta }(\PPP)
:= {\GGG}^{loc}_{\eta\bar\eta }(\PPP)\Big|_{\pss\ge s_1+s_2} \,
\q\GGG^{<\,loc}_{\eta\bar\eta }(\PPP):= {\GGG}^{loc}_{\eta\bar\eta }(\PPP)\Big|_{\pss<s_1+s_2} \,
\ee
 are projections of $\GGG^{loc}_{\eta\bar\eta }$
to the respective regions of spins.

To find canonical gauge invariant current deformation, that admits a proper flat limit, let us
 decompose $\GGG^{\ge\,loc}_{\eta\bar\eta }$ \eqref{tworegloc}
 as a sum of eigenvectors
of two mutually commuting operators
 $\hat{s}$ \eq{hats} and $f_0$ \eq{slv}
\ie
 \bee\label{seriesDEF} \GGG^{\ge\,loc}_{\eta\bar\eta }(\PPP) =\frac{i }{8} \eta\bar\eta
\sum_{\pss\ge1}\,\,\sum_{ -\pss\le \hf \le \pss } \GGG^{ loc} (\PPP)\big|_{\pss, \hf}
\,.\eee
where
\bee \nn\hat{s}\,\GGG^{ loc} (\PPP)\big|_{\pss, \hf}&=&2  (\pss-1) \GGG^{ loc} (\PPP)\big|_{\pss, \hf}
 \\ \nn f_0 \,\GGG^{ loc} (\PPP)\big|_{\pss, \hf}&=&2  \hf \GGG^{ loc} (\PPP)\big|_{\pss, \hf}
 \,.\eee

To this end, using \eq{slv}  along with the fact that
 $y_\ga y^\ga=\by_\pa \by^\pa=0$, it is convenient   to represent
$\GGG^{\ge\,loc}_{\eta\bar\eta }$ \eqref{DXloc}
in a slightly different form resulting from the substitution
\be\nn \p_1{}_{\ga}\p_2{}^{\ga}\to f_-
\q\bp_1{}_{\pa}\bp_2{}^{\pa} \to -f_+\,, \ee
 where, introducing
 the contour integrations   over cycles close to zero,
\bee   \label{seriesDEFsn}\ls
   \GGG^{ loc}\!(J) \big|_{\pss, \hf}\!\!\! &=&  \!\!\! \frac{i}{8}
\int       \f{ d^4\gt}{  \gt_4^2 }    \gd(1\!-\!\gt_3\!-\!\gt_4)\gd'(1\!-\!\gt_1\!-\!\gt_2)
\Upsilon
\oint \f{dw}{2\pi i w^{2\pss+1}}
\oint \f{dv}{2\pi i v^{2\hf+1}}
  \qquad\\ \nn&&
  \left\{\rule{0pt}{14pt}\right. \bar{H}^{\pa\pb}
\bp_\pa\bp_\pb   %
  \exp (\!-\!i   \gt_3 f_+
 v^{ 2}  )  \PPP(    \gt_1y  v w,  \!-\!  \gt_2 y  v w,   \gt_4 \gt_2\by  v^{\!-\!1} w,
\!-\!\gt_4 \gt_1\by v^{\!-\!1} w;K)
 \qquad
\qquad\\ \nn &+&    H^{\ga\gb}
\p_\ga\p_\gb   %
        \exp( i   \gt_3f_-
         v^{\!- 2})
    \PPP(   \gt_4\gt_1y v w ,  \!-\! \gt_4 \gt_2 y  v  w,    \gt_2\by v^{\!-\!1} w,
  \!-\!\gt_1\by v^{\!-\!1} w;K) \left.\rule{0pt}{14pt}\right\} .
\qquad\eee
 For any $|\hf|\le\pss $,  $ \GGG^{ loc}  (\PPP  )\big|_{\pss, \hf}$
 is a   consistent deformation, that is gauge invariant
 since inequality \eq{sss} holds by construction.

So defined  $ \GGG^{ loc}  (\PPP  )\big|_{\pss, \hf}$  describes
 the spin-$\pss$ contribution of $\PPP_\hf $ in the one-form sector
 since  the total degree  in $ y$ and $\by$ is
 $2(\pss-1)$.
Moreover, $\GGG_{\eta\bar\eta}^{ loc} (\PPP )\big|_{\pss, \hf}  $
projects currents to the components $\PPP_\hf$  obeying   $f_0 \PPP_\hf =2\hf \PPP_\hf $, \ie
  $\GGG^{ loc}  (\PPP )\big|_{\pss, \hf} \equiv
  \GGG^{ loc}  (\PPP_\hf )\big|_{\pss, \hf} .$

 However,
 only   deformations $\GGG^{\,loc} (\PPP_\hf )\big|_{\pss, \hf}$ with   $|\hf |\le \half$   admit
  a proper flat limit (for more detail see  \cite{Gelfond:2010pm}).
 On the other hand,
  using  analogues of the manifest formulae for  trivial deformations
  of \cite{Gelfond:2010pm} it will be shown that,  up to a numerical coefficient,
      a    current deformation  $\GGG^{\,loc} (\PPP_\hf )\big|_{\pss, \hf}$
  at  $\pss\ge2$, $ \pss\ge \hf >\half$ in $AdS_4$ is equivalent  modulo
 a local field redefinition to $\GGG^{\,loc} ( (f_-)^\hf  \PPP_\hf )\big|_{\pss, 0}$  for integer $\hf $
 and    $\GGG^{\,loc} ( (f_-)^{\hf -\half} \PPP_\hf )\big|_{\pss, \half}$   for half-integer $\hf $.
  Analogously, a current  deformation   associated with   $\PPP_{ \hf }$    at
   $-\half>\hf\ge-\pss $ is equivalent up to a numerical coefficient to $\GGG^{\,loc} ( (f_+)^{-\hf}  \PPP_\hf )\big|_{\pss, 0}$  for integer $\hf $ and
 $\GGG^{\,loc} ( (f_+)^{-(\hf+\half)}  \PPP_\hf )\big|_{\pss, -\half}$
 for  half-integer $\hf $.

   Thus the proper strategy for reducing a local current interaction to the canonical form
    that admits  flat limit is to add improvements involving
   $f_\pm$  to remove all contributions of currents with $|\hf|> \half$ to achieve
   that the resulting deformation  would only  involve {\emph {canonical currents}}
   $\widetilde{\PPP}_{ {\bf m }}\sim(f_\pm)^{ [|\hf|]}  \PPP_\hf $ with $  {\bf m }=\pm \half $ or $0$.

Note that this procedure simultaneously removes   contributions
 to the \rhs of the HS torsion-like tensor for integer spins, proportional to   $  y^{ s-1}\by^{ s-1}$.
Indeed,    $ \widetilde{\PPP}_{\bf0}$   does not contribute to
the  torsion-like terms
    because of the pre-factors   $\overline H^{\pa\pb}  \bp_i{}_\pa\bp_j{}_\pb$ { and  }
 $H^{\ga\gb}  \p_i{}_\ga\p_j{}_\gb$ in  (\ref{seriesDEFsn}). Hence,
 \emph {canonical currents} do not contribute to torsion.


Now we are in a position to explain details of the canonical current
construction.

\subsubsection{Canonical currents}
\label{Canonical deformations2}
Our aim is to find such local one-form   $\Lambda^{sum}$
and zero-form  $B^{sum}$    that the field redefinition \eq{redifcan}
 reduces the current interactions to the canonical  form \eq{canonical ressub==}.

Let $\PPP$  be a solution  to  current equation \eq{tw2}.
 Consider  the  one-form
\bee\label{Omeganewexp}\Lambda(\PPP) =  \frac{i }{8} \eta\bar\eta  \,h(\p,\bp)
\int_0^1 \f{d\gt_3 }{1-\gt_3 }
        \int d^2\gt   \gd'(1\!-\!\gt_1\!-\!\gt_2)
   \theta(\gt_1)  \theta(\gt_2) \,
      \qquad\\ \nn   \exp (i \gt_3\,f_-)
        \, \PPP  (   (1-\gt_3)\gt_1y+y^1,  - (1-\gt_3) \gt_2 y+y^2 \,,
         \gt_2\by+\by^1, -\gt_1\by+\by^2;K|x)\,\big|_{ y^j=\by^j=0}\,
 \eee
 analogous to the form $\Omega$ introduced in Appendix $D$ of \cite{Gelfond:2010pm}
 for a similar  purpose.
(Note that the fictitious  pole in $1-\tau_3$ is  compensated by the $y$-differentiation
$\p$ in $h(\p,\bp)$.)  Differentiation of  $\Lambda(\PPP)$ gives
upon  $\gt_3$-integration
in the $ \bar{H}$-dependent term
\bee\label{OmeganewexpD}\D_{ad}\Lambda (\PPP)=
 -\frac{i }{8} \eta\bar\eta \int d^2\gt   \gd'(1\!-\!\gt_1\!-\!\gt_2)
   \theta(\gt_1) \, \theta(\gt_2) \, \,
    \qquad \\ \nn
    \Big\{
 \bar{H}^{\pa\pb}\bp_\pa\bp_\pb  \Big(
       \exp(i    {f}_-)
        \, \PPP (   y^1,   y^2  ,  \gt_2\by+\by^1, -\gt_1\by+\by^2;K|x)
   \qquad \\ \nn
     -   \PPP (    \gt_1y+y^1,  -   \gt_2 y+y^2  ,  \gt_2\by+\by^1, -\gt_1\by+\by^2;K|x)
   \Big)    \qquad
  \\ \nn
-H^{\ga\gb}\p_\ga\p_\gb \int_0^1 \f{d\gt_3}{(1-\gt_3)^{ 2}}
          \,\,
 \big[-(1-\gt_3)\by^\pa\bp_\pa+i {f}_+
\big]   \exp(i \gt_3\,\widetilde{f}_-)\qquad\\ \nn
        \, \PPP \ (   (1-\gt_3)\gt_1y+y^1,  - (1-\gt_3) \gt_2 y +y^2\,,   \gt_2\by+\by^1, -\gt_1\by+\by^2;K|x)
        \Big\}\,\big|_{ y^j=\by^j=0}\,,
\eee
where $$\widetilde{f}_-:=  ( \gt_2\by+\by^1)_\pa( -\gt_1\by+\by^2)^\pa +\p_1{}_\ga\p_2{}^\ga.$$

Taking into account  Eqs.~(\ref{znakH})-(\ref{dlor}), \eq{commfkf} and using
decomposition $$\Lambda (\PPP)=\frac{i }{8} \eta\bar\eta\sum_{\pss\ge1} \sum_{|\hf|\le\pss}\Lambda (\PPP)|_{\pss,\hf}\q$$
 \bee\label{Omeganewexpsn}
 \ls
  \Lambda(\PPP )\big|_{\pss,\hf} =
 \,h(\p,\bp)
\int_0^1 \f{d\gt_3 }{1-\gt_3 }
        \int d^2\gt   \gd'(1\!-\!\gt_1\!-\!\gt_2)
 \Upsilon \oint \f{dw}{2\pi i w^{2\pss+1}}
\oint \f{dv}{2\pi i v^{2\hf+1}}
        \exp( i   \gt_3f_-  )%
  \qquad\\ \nn
     \PPP(  (1- \gt_3)\gt_1y v w \!+\!vy^1,  \!-\! (1-\gt_3) \gt_2 y  v  w\!+\!vy^2,    \gt_2\by v^{\!-\!1} w,  
  \!-\!\gt_1\by v^{\!-\!1} w ;K) \big|_{ y^j=\by^j=0} ,
 \eee
 straightforward   computation analogous to that of Appendix $D$ of \cite{Gelfond:2010pm}
 yields
 \be \label{DOMEGAbH4DEL}
  \D_{ad}{\Lambda}  (\PPP ) \big|_{\pss,\hf} = (\pss- \hf -1) \GGG^{\ge\,loc}_{\eta\bar\eta}
  (\PPP )\Big|_{\pss,\hf }
  - \GGG^{\ge\,loc}_{\eta\bar\eta} ([-if_+\PPP ] ) {\big|_{\pss,\hf+1} }
+ \frac{i }{8} \eta\bar\eta\bar{H}^{\pa\pb} \bp_\pa  \bp_\pb
B(  \PPP)\big|_{\pss,\hf }
\,\qquad
  \ee with  $\GGG^{\ge\,loc}_{\eta\bar\eta} (\PPP)\Big|_{\pss,\hf }$  \eq{seriesDEFsn} and
  \bee \label{Krajsn}
 {B}(\PPP)\big|_{\pss,   \hf }=
-         \int d^2\gt   \gd'(1\!-\!\gt_1\!-\!\gt_2)
 \Upsilon \oint \f{dw}{2\pi i w^{2\pss+1}}
\oint \f{dv}{2\pi i v^{2\hf +1}}
  \qquad\\ \nn
        \exp(i   f_-  )
        \, \PPP  (    vy^1,   vy^2   ,  \gt_2\by w v^{\!-\!1}\! -\gt_1\by w v^{\!-\!1};K|x)
    \big|_{ y^j=\by^j=0} \,.    \qquad         \eee

By virtue of \eq{DOMEGAbH4DEL}, for $\pss-\hf>1$,   the field redefinition
\be\label{f+redef} \go \to  \go+\frac{i }{8} \eta\bar\eta\f{1}{(\pss - \hf  -1)}
\Lambda (\PPP )\big|_{\pss,\hf}\q
C \to   C +  \frac{1 }{2}  \bar\eta\f{1}{(\pss-\hf  -1)} {B}(\PPP)\big|_{\pss,{   \hf}}
 \q
 \ee   replaces $\GGG^{\ge\,loc}_{\eta\bar\eta }(\PPP )\Big|_{\pss,\hf }$ in Eq.~\eq{W2hh'} by
 \be\label{onestepsdvig}
  \f{1}{(\pss  -\hf  -1)} \GGG^{\ge\,loc}_{\eta\bar\eta }( -if_+\PPP )\Big|_{\pss,\hf+1 }\,.\ee

Since $ f_+  \PPP  $
again satisfies  current equation, this procedure can be repeated.
For $-\pss\le\hf\le-1  $, to bring the current to the canonical form one needs
 $k$ steps with $k=[|\hf|]\equiv|\hf|-\{|\hf|\}$.

 For  any $\pss $, the field redefinition
  \be \label{f+redef1} \go \to\go+\Lambda^-({\pss,\hf }) \q
   C \to C+ {B} ^-({\pss,\hf }) \ee
   with
   \bee\label{Lambd-}  \Lambda^-({\pss,\hf }) &=&\frac{i }{8} \eta\bar\eta
   \sum_{k=1}^{[|\hf|]} \f{(\pss +|\hf|  -k-1)!}{(\pss +|\hf|  -1)!}
 \Lambda ((-if_+)^{k-1}\PPP_{-|\hf|} )\big|_{\pss,-|\hf|+k-1}
 \q
\\ {B} ^-({\pss,\hf })&=& \half\bar \eta \sum_{k=1}^{[|\hf|]} \f{(\pss +|\hf|  -k-1)!}{(\pss +|\hf|  -1)!}{B}
((-if_+)^{k-1}\PPP_{-|\hf|} )\big|_{\pss,-|\hf|+k-1}
\eee
gives the canonical current in the form
 \be
  {\GGG }^-({\pss,\hf }) =  \f{(\pss-1 +\{|\hf|\} )!}{(\pss  +|\hf|   -1)! }
  \GGG^{\ge\,loc}_{\eta\bar\eta } (  (-if_+)^{[|\hf|] }\PPP )\big|_{\pss ,-\{|\hf|\}  }\,. \qquad
\ee

 The generating functions  for $\Lambda^-({\pss,\hf }) $ and  ${B}^-({\pss,\hf }) $
with $\pss$ and $\hf $ satisfying  $-\pss\le\hf\le-1$ are
   \bee\label{OmegaFRgen}
   \Lambda^-= \frac{i }{8} \eta\bar\eta   \oint \f{dv v}{2\pi i (1-v) }
\int \f{ d^4\gt d^2\gr}{\gr_1^2\gt_4}\gd(1\!-\!\gr_1\!-\!\gr_2)
   \gd'(1\!-\!\gt_1\!-\!\gt_2)\gd(1\!-\!\gr_3\!-\!\gr_4)
\Upsilon
\qquad \\ \nn
h(\p,\bp)\exp   (i \gr_1^{-1}\gt_3 f_-)
   \exp ( i\gr_2  v^2( (  \gt_4\gt_1y_\ga+y^1{}_\ga)
         (\!-\! \gt_4\gt_2 y^\ga +y^2{}^\ga\,) +\bp_1{}_\pa\bp_2{}^\pa) ) \quad\\ \nn
   \PPP(    (\gt_4\gt_1y   \!+\!  y^1)v,  (\!-\!\gt_4 \gt_2 y   \!+\!  y^2)v ,(\gr_1
  \gt_2\by \!+\! \by^1)v^{\!-\!1}  ,(
\!-\! \gr_1 \gt_1\by   \!+\! \by^2)v^{\!-\!1} ;K)  \big|_{ y^j=\by^j=0} \,,
  \quad  \eee
 \bee \label{Krajsnsubgen}
 {B}^-=\frac{1 }{2}  \bar\eta \oint \f{dv v}{2\pi i (1-v) }\int  d^2 \gt\f{d^2\gr}{\gr_1^2}\gd(1-\gr_1-\gr_2)
    \Upsilon    \gd'(1\!-\!\gt_1\!-\!\gt_2)
\rule{100pt}{0pt} \\ \nn
 \exp(  i  \gr_1^{-1} f_ -\!\big)  \exp (-i\gr_2 {f}_+ \, v^2 ) \PPP( y^1 v , y^2 v ,
  (\gr_1\gt_2\by  \!+\!  \by^1)v^{\!-\!1}  ,
(\!-\!  \gr_1\gt_1\by   \!+\! \by^2)v^{\!-\!1} ;K)
  \big|_{ y^j=\by^j=0} \,,
       \qquad
         \eee
        where    the measure
         $\f{dv v}{2\pi i (1-v) }\equiv \f{dv }{2\pi i v  }(v^2+v^3+\ldots)$
implies   summation over such $\PPP$
 that $f_0$-eigenvalue of $  f_+  \PPP$ is    smaller than $-1$.
 One can see, that negative degrees in $\gr_1$ in \eq{OmegaFRgen} and \eq{Krajsnsubgen}  do not
  survive upon  integration over $v$.

 The resulting canonical current
         is
 \bee\label{kstepsdvigres}
      \GGG ^{-} \,\ls&=&  \!\! \f{1}{8}\eta\bar\eta \oint \f{dw}{2\pi i w }
      w  ^{- 2|\pss|_2}\int \f{d^2\gr}{\gr_1 }\f{ d^4\gt}{ \gt_4^{2}}
     \gd'(1-\gr_1-\gr_2)
   \gd (1-\gt_3-\gt_4)\gd'(1\!-\!\gt_1\!-\!\gt_2) \qquad \\ \nn &&\Upsilon
  \left[\rule{0pt}{12pt}\right.
 \bar H^{\pa\pb}   {\bp_\pa \bp_\pb}
 \exp  \big (- i (\gr_1\gt_3 +w^{\!-\!2}\gr_2)f_+)
 \qquad\\ \nn&&\PPP(    (\gt_1y  \!+\!   y^1) w^{\!-\!1} ,
 (\!-\!   \gt_2 y \!+\!   y^2)  w^{\!-\!1}  ,
  (\gr_1\gt_4\gt_2\by  \!+\!  \by^1) w   ,(
\!-\! \gr_1\gt_4 \gt_1\by \!+\!  \by^2) w  ;K)
\rule{25pt}{0pt}
\\ \nn &&+   H^{\ga\gb} {\p_\ga \p_\gb}(\gr_1)^{ 2|\pss|_2}
   \exp\big ( i  \gr_1 \gt_3f_-\big)
   \exp(-i\gr_2\widetilde{f}_+ w^{\!-\!2})
  \\ \nn &&\PPP(  (  \gr_1\gt_4 \gt_1y  \!+\!   y^1) w^{\!-\!1} ,
  ( \!-\! \gr_1\gt_4 \gt_2 y   \!+\!   y^2)w^{\!-\!1}  ,
  (\gt_2\by   \!+\!  \by^1)w   ,
(\!-\!   \gt_1\by \!+\!  \by^2) w   ;K) \left.\rule{0pt}{12pt}\right]
 \Big|_{ y^j=\by^j=0}\,,
\eee
  where
  \be\label{f+ot2}\widetilde{f}_+:=  ( \gr_1 \gt_4  \gt_1y_\ga+y^1{}_\ga)
         (  \gr_1\gt_4  \gt_2 y^\ga -y^2{}^\ga\,) -\bp_1{}_\pa\bp_2{}^\pa \,.\ee
 The measure
 $\dis{\f{dw}{ 2\pi i w}
    w^ { -|2\pss|_{ 2}} }$
  is  different for bosonic currents with integer $\pss$ ($|2\pss|_{ 2}=0$) and
fermionic ones with half-integer $\pss$ ($|2\pss|_{ 2}=  1$).\rule{0 pt}{15pt}

  The complex conjugated case with  $\overline{-if_+}= i f_-  $ is analogous.
   The respective generating functions
    can be obtained from \eq{OmegaFRgen}-\eq{kstepsdvigres} via the replacement $$
    i f_-\leftrightarrow -if_+\q y\leftrightarrow\by\q H \leftrightarrow \bar H\q
     $$ that, abusing terminology, will be referred to as $c.c.$ though the sign of the
      overall factor of   $i$ is  not changed.

 Summarizing,   by virtue of  Eq.~\eq{DOMEGAbH4DEL},    local field redefinition  \eq{redifcan}  with
      \bee \label{Omeganewexpgen}&&\Lambda^{sum}(\PPP) =
 \Lambda^- +c.c.  \,,\\&&
      \label{sdvigSint}
       {B}^{sum} (\PPP) = \bar{B}^- +c.c.\q \eee
       with
 $\Lambda^-$ \eq{OmegaFRgen}, $\bar{B}^-$ \eq{Krajsnsubgen}
    leads to
   deformed equations  \eq{W2hhloc0}, \eq{W2hhloc0tw}. More precisely,
    taking into account that both $ {\GGG}^{-}$ and its complex conjugate contain the same
    $\gr_2$-independent term,  to avoid the double counting, we add to $ \GGG ^{-} (\PPP)+c.c.$
    the term proportional to $ \gd(\gr_2)$ obtaining $ \GGG ^{can} (\PPP) $ in the form
    \eq{canonical ressub==} upon the substitution of $f_\pm$ \eq{slv} and
    reformulation of the final result in the form of a Gaussian integral.

Note that there is a freedom in the field redefinition in the zero-form sector, that does not affect
corrections to dynamical field equations in the one-form sector.
Indeed, addition to $   {B}^{sum} (\PPP)$ \eqref{sdvigSint} of any 
field containing a factor  of $ y\by $  does not affect Eq.~\eq{CON1k}.
For instance, for $\pss=1$, $\hf=\pm1$
it is convenient to use a  field redefinition of this type
to obtain the conventional form of the Maxwell equations as discussed
in Section \ref{s=1}.

\section{Current contribution to dynamical equations }
\label{contribution}
In this section we derive explicit form of the current contribution to the
 {\it r.h.s.} of  massless equations of different spins that follows from the
 nonlinear HS equations. Note that the results of this section
  extend the variety of examples of current interactions explicitly
presented in \cite{Gelfond:2010pm} to all sets of spins respecting inequality (\ref{sss}).

For the future convenience we will use the following decompositions
 \bee\label{jdecom}
  A(y_{1,2},\by_{1,2}| x)= \sum_{m,\bm } A_{m\,,\bm}(y_{1,2},\by_{1,2}| x)\q
     B(y ,\by | x)= \sum_{m,\bm } B_{m\,,\bm}(y,\by| x)\,,
  \eee
  with
  \be\nn
\big( y^1{}^\gb  \p_1{}_\gb + y^2{}^\gb \p_2{}_\gb  \big)
 A_{m\,,\bm}(y_{1,2},\by_{1,2}| x)=m A_{m\,,\bm}(y_{1,2},\by_{1,2}| x)\,,
\ee
\be \nn
\big( \by^1{}^\pb \bp_1{}_\pb +\by^2{}^\pb \bp_2{}_\pb\big)
A_{m\,,\bm}(y_{1,2},\by_{1,2}| x)=\bm A_{m\,,\bm}(y_{1,2},\by_{1,2}| x)\,,
\ee
\be\nn y^\gb \p {}_\gb   B_{m\,,\bm}(y,\by| x)=m B_{m\,,\bm}(y,\by| x)
 \q
 \by^\pb \bp {}_\pb    B_{m\,,\bm}(y,\by| x)=\bm B_{m\,,\bm}(y,\by| x)\,.\ee
Recall, that we consider currents \eq{Csumkbark'} that are bilinear in
fields represented as series \eq{spinsC}   in $y$ and $\by$.

  \subsection{Spin $ 0$}

As mentioned in Section \ref{Currinter}, gauge invariant
 currents $J$ of spin zero are built from the zero-forms $C$ carrying $s_1=s_2=\half $
(the  $\pss=0$ and $\pss= \half$
 conformal currents are not conserved since the fields of spins
$\pss=0$ and $\pss= \half$ are not gauge).

  By virtue of \eqref{tw}, \eqref{hh} and  \eqref{barhh}, taking into account  \eq{sdvigSint} and \eq{OmegaFRgen},
 Eq.~(\ref{W2hhloc0tw}) yields for $\PPP= \PPP_{\bf \pm1}$
  \be \label{CON21j}
  D^{L}{}_{\ga\pa}C( K| x) +i
  C_{\ga\pa}( K| x) =0\,\,,\qquad\qquad\qquad\qquad\ee
  \bee&{}&
  \rule{0pt}{22pt} \nn D^{L}{}_{\ga\pa}C_{\gb\pb}( K| x)  +i
  C_{\ga\gb\pa\pb}( K| x) -i\gl \gvep_{\pa\pb}\gvep_{\ga\gb}C( K| x)  \qquad\\
  &{}&
 \nn - \frac{i}{4}{\bar \eta}   \exp i [  \p_1{}_\gb   \p_2{}^\gb ]
   {\gvep_{\pa\pb}} (\p_1{}_\gb+ \p_2{}_\gb)
    (   \p_2{}_\ga -   \p_1{}_{\ga} )
   \PPP_{\bf 1} (  y^1 ,  y^2 ; \bar  y^1   , \bar  y^2  ;K|x) *\bar k\,\big|_{ y^j=\by^j=0}
   \\ &{}&\nn -\frac{i}{4}\eta
  \exp{(i [ \bp_1{}_\dgb  \bp_2{}^\dgb ])}
 { \gvep_{\ga\gb}}  (\bp_1{}_\pb+ \bp_2{}_\pb)(\bp_2{}_{\pa} -  \bp_1{}_{\pa})
  {\PPP}_{-\bf 1} (  y^1 ,  y^2 ; \bar  y^1   , \bar  y^2  ;K|x) *k\,\big|_{ y^j=\by^j=0}
    =0.\qquad
\eee
Contracting indices one obtains  by virtue of \eq{Csumkbark'} that the respective contribution of the currents
$\PPP_{\bf \pm1}$ bilinear in the fields $C$     with   $ s_1=s_2=\half$
  is
 \bee\label{CONbilindoubl=}
   D^{L}{}^{\ga\pa} D^{L}{}_{\ga\pa}  \sum_{j=0,1} C^{j,1-j}( x) k^j\bar k{}^{1-j}
 +   \bar\eta
    \sum_{j,\,l=0,1} (-1)^j C {}_\ga^{j,1-j}( x) 
    C{}^{l,1-l}{\,}^\ga( x)  k^{l+j}\bar k{}^{  1-l-j}  \,
  \qquad\\ \nn
  + \eta
 \sum_{j,\,l=0,1}(-1)^{1-j} C{}_\pa^{j,1-j}( x)
  C{}^{l,1-l}{\,}^\pa( x)     k^{1+l+j}\bar k{}^{-l-j}\,
     = 0\qquad
   \eee
 just
reproducing  Yukawa interaction since $C{}_\ga ( x)$ and
$ {C} {}_\pa(  x )$ are dynamical spin-$1/2$
fields.  Note that a $C^2$-deformation, that one might naively expect in the
spin-zero  sector, does not appear in agreement with the fact shown in
\cite{Gelfond:2015poa} that the interactions with gauge invariant currents
are conformal in $d=4$, while the $C^2$-deformation is not.

   \subsection{Spin   $ 1/2$}
\label{exampshalfj}

  By virtue of \eqref{tw},  taking into account \eq{hh}, \eq{sdvigSint} and \eq{OmegaFRgen}
 Eq.~(\ref{W2hhloc0tw}) yields    \be \label{CON22j1/2}
\!\!D^{L}{}_{\ga\pa}C_\gga( K| x) +i
C_{\gga\ga\pa}( K| x) + \f{\eta}{4} \gvep_{\gga\ga}
   ( \bp_1{}_{\pa} \!-\! \bp_2{}_{\pa})\exp{(i [ \bp_1{}_\dgb  \bp_2^\dgb ])}
\PPP_{ }( K| x) *k\big|_{y^j=\by^j=0}\,
=0 .\ee
 Hence
\be\label{dirj}
D^{L}{}_{\ga\pa}C^\ga ( K| x) \!-   \f{\eta}{2}
     ( \bp_1{}_{\pa}\! -\! \bp_2{}_{\pa})\exp{(i [ \bp_1{}_\dgb  \bp_2{}^\dgb ])}
\PPP ( K| x) *k\big|_{ y^j=\by^j=0}\,
 =0. \ee
Substitution of   bilinear currents
  $\PPP_{\bf-\half}\,$   \eq{Csumkbark'}  built   from  the fields of spins 0 and 1/2  gives
 the Yukawa interaction in the spin-$1/2$ sector
   \bee\label{dirjbe}\!\!
 D^{L}{}_{\ga\pa}\sum_{j=0,1} C^{j,1-j}{}{\,}^\ga(   x)k^j\bar k{}^{1-j}  -\!   \half\eta
 \sum_{j,\,l=0,1} C^{j,1-j}{}{}_\pa( x )
  C^{l,1-l}{}  ( x ) k^{1+l+j}\bar k{}^{ -l-j}
\quad\\ \nn\! +  \half \eta  \sum_{j,\,l=0,1} (-1)^{1-j}
 C^{j,1-j}{} ( x  )
  C^{l,1-l}{}{}_\pa( x ) k^{ l+j}\bar k{}^{1-l-j}  =0
\,.
\qquad\eee
Analogously,
\bee\label{dirjbecc}\!\!
 D^{L}{}_{\ga\pa}\sum_{j=0,1} C^{j,1-j}{}{}^\pa(  x)k^j\bar k{}^{1-j}  -\!   \half\bar\eta
 \sum_{j,\,l=0,1} C^{j,1-j}{}{}_\ga( x )  C^{l,1-l}{}  ( x ) k^{1+l+j}\bar k{}^{-l-j}
\quad\\ \nn\! +  \half \bar\eta  \sum_{j,\,l=0,1} (-1)^{ j}
 C^{j,1-j}{} ( x  )
   C^{l,1-l}{}{}_\ga( x ) k^{ l+j}\bar k{}^{1-l-j}  =0
\,.
\qquad\eee

\subsection{Maxwell equations}
  \label{s=1}


   Eq.\eqref{W2hhloc0}   
       still reads as (\ref{CON1k})
\be \label{CON11}
 \!\!\mathcal{D}_{ad}\omega(0,0;K|x)  =
   \f{ i}{4} \Big ( \eta \bar{H}^{\dga\pb}\bp_{\pa} \bp_{\pb}
{  C }(0,\by;K|x)*k  +\bar \eta H^{\ga\gb} \p_{\ga} \p_{\gb}
{C }(y,0;K|x)*\bar k\Big )\big|_{y=\by=0}
.\quad\ee
This identifies $\bar \eta C_{\ga\gb}(x)$ and $\eta \overline{C}_{\pa\pb}( x)$,
respectively,
with the self-dual and anti-self-dual parts of the Maxwell field strength.

Using a freedom in local field redefinitions in the zero-form sector
 mentioned in Section \ref{Canonical deformations2}
 it is convenient to change ${ B}^{sum}$ (\ref{sdvigSint}) to ${ \widetilde B}^{sum}$
\be\label{phi1} { B}^{sum}\to \widetilde{ B}^{sum}= { B}'   (\PPP_{\bf 1})
+ \overline{B}'   (\PPP_{\bf -1})\q \ee where\bee \nn
 { B}'   (\PPP_{\bf 1})&=& \half \eta \int    d^2\gt   \gd'(1\!-\!\gt_1\!-\!\gt_2)
 \PPP_{\bf 1}(\tau_1 y, - \tau_2 y , \bar y  , \bar y ; K)*\bar k\,,
\\\nn\overline{ B}'    (\PPP_{\bf -1})&=&\half \bar\eta\int    d^2\gt   \gd'(1\!-\!\gt_1\!-\!\gt_2)  \PPP_{\bf -1}
( y, - y , \tau_1\bar y   , \tau_2 \bar y ; K)*k
\, .
\eee
Recall, that  $ \hf$ in $\PPP_\hf$ (\ref{bilinhfCC})
is the   sum of helicities  of the constituent fields. Also note that
 the additional $\PPP_{\pm 1}$-dependent local field redefinition (\ref{phi1}) was not
discussed in \cite{Vasiliev:2016xui} where the contribution
of currents $\PPP_{\pm 1}$ was not considered.

Evidently, by virtue of   \eq{sdvigSint} and \eqref{phi1}\be
H^{\ga\gb}\! \p_{\ga} \p_{\gb}
{ B}'  (\PPP_{\bf 1})(y,0;K| x)\equiv 4    H^{\ga\gb}\!\p_{\ga} \p_{\gb}
B^{sum}(\PPP_{\bf_1})(y,0;K| x) \,,\quad
 \ee\be
\bar{H}^{\pga\pb} \bp_{\pga} \bp_{\pb}
\overline { B}'  (\PPP_{\bf -1})(0,\bar y;K| x)\equiv 4
\bar{H}^{\pga\pb} \bp_{\pga} \bp_{\pb}
 B^{sum}(\PPP_{\bf -1})(0,\bar y;K| x)\,.\quad\ee
Hence, in the $\pss=1$ sector  Eq.~\eqref{W2hhloc0tw}  is equivalent to
\bee \label{HHloc1}
 \D_{tw} C  &=& -   \Hhh^{loc}_\eta{}_{cur} (\PPP_{\bf -1})  \,-  \Hhh^{loc}_{\bar{\eta}}{}_{cur} (\PPP_{\bf 1}) \,
 + \D_{tw}  { B}'   (\PPP_{\bf 1})
+ \D_{tw}\overline{B}'  (\PPP_{\bf -1})\,.
  \eee
The following useful formula results from \eq{phi1} by the straightforward computation
 \bee
\label{Dtwphi2}
 D_{tw} { B}'  (\PPP_{\bf 1})= \f{i}{2} \eta    h^{\ga\pa} \int  {  d^2\gt}
 \, \gd (1-\gt_1-\gt_2) \Big(
-\gt_1 y{}_{\ga} \bp_2{}_{\pa} +\gt_2 y{}_{\ga} \bp_1{}_{\pa} \Big)
\\ \nn
 \{\p_1{}_\gga\p_2{}^\gga\PPP\}(\gt_1 y , -\gt_2 y    ; \gt_2 \by  , -\gt_1 \by ; K)*k
\,.\qquad \eee

 Deformed  equation \eqref{HHloc1}   in the $\pss=1$ sector
yields by virtue of Eqs.~\eqref{hh}, \eqref{barhh}, \eqref{phi1}   and
\eq{Dtwphi2}
   \bee\label{CON22j}
D^{L}{}_{\ga\pa}C_{\gb\gga}(K| x)\, +\,i
C_{\gb\gga\ga\pa}( K| x)
 - \frac{ 1}{2}\eta \big[\gvep_{\gb\ga}  \p_\gga+\gvep_{\gga\ga}  \p_\gb\big]
  \int_0^1 d\tau  ( \tau\bp_2{}_{\pa} - (1-\tau) \bp_1{}_{\pa})
\qquad \\ \nn\Big\{
      \PPP_{\bf 0}+i   \bp_1{}_\dgb  \bp_2{}^\dgb     \PPP_{\bf - 1}
 -i \p_1{}_\gb\p_2{}^\gb \PPP_{\bf 1}
 \Big\}
(\tau y , -(1-\tau) y , \bar y   , \bar y ;K|x) *k\,\big|_{ y=\by=0} \, =0\,.  \eee
 Contracting indices we obtain from \eq{CON22j}
 \bee\label{CON22jc}
D^{L}{}^{\gb}{}_{\pa}C_{\gb\gga}(0\,,0 ;K| x)\,
 + \frac{ 3}{2}\eta
  \int_0^1 d\tau   (\gt\p_1{}_\gga-(1-\gt)\p_2{}_\gga) ( \tau\bp_2{}_{\pa} - (1-\tau) \bp_1{}_{\pa})
\qquad \\ \nn\Big\{\PPP_{\bf 0}+i   \bp_1{}_\dgb  \bp_2{}^\dgb     \PPP_{\bf - 1}
 -i \p_1{}_\gb\p_2{}^\gb \PPP_{\bf 1}
  \Big\}
   (  y^1 ,  y^2 ; \bar  y^1   , \bar  y^2  ;K|x) *k\,\big|_{ y^j=\by^j=0} \, =0\,. \eee
Analogously,
\bee\label{CON22jcc}
D^{L}{}_{\gga}{}^{\pb}C_{\pb\pa}(0\,,0;K| x)\,
 - \frac{ 3}{2}\bar\eta
  \int_0^1 d\tau   (\gt\bp_1{}_\pa -(1-\gt)\bp_2{}_\pa) ( \tau\p_2{}_{\gga} - (1-\tau) \p_1{}_{\gga})
\qquad \\ \nn\Big\{\PPP_{\bf 0} +i \p_1{}_\gb\p_2{}^\gb \PPP_{\bf 1}
-i   \bp_1{}_\dgb  \bp_2{}^\dgb     \PPP_{\bf - 1}
  \Big\}
    (  y^1 ,  y^2 ; \bar  y^1   , \bar  y^2  ;K|x) *\bar k\,\big|_{ y^j=\by^j=0} \, =0\,.  \eee
 Hence, performing integration over $\tau$, from \eq{CON22jc} and \eq{CON22jcc}
 it follows  for $D^{L}=h^{\gga\pga}D^{L}{}_{\gga\pga}$ that 
 \bee\label{Maxsum}
 \eta D^{L}{}_{\gga}{}^{\pb}C_{\pb\pa}(0\,,0;K| x)*\bar k+
 \bar\eta D^{L}{}^{\gb}{}_{\pa}C_{\gb\gga}(0\,,0;K| x)\,
 * k\qquad \\ \nn=  \half \eta\bar\eta(
      2\bp_1{}_\pa \p_2{}_{\gga}- \bp_2{}_\pa  \p_2{}_{\gga}
      - \bp_1{}_\pa  \p_1{}_{\gga}+2\bp_2{}_\pa   \p_1{}_{\gga})
 \PPP_{\bf 0}
    (  y^1 ,  y^2 ; \bar  y^1   , \bar  y^2  ;K|x)  \,\big|_{ y^j=\by^j=0} \, .
\eee
Using  identities
\be\label{iden2-3}
H^{\ga\gb}  h^{\gga \pga} = \epsilon^{\ga\gga} \Hhh^{\gb\pga}+
\epsilon^{\gb\gga} \Hhh^{\ga\pga}\q
\bar{H}^{\pa\pb} h^{\gga \pga} =
-\epsilon^{\pa\pga} \Hhh^{\gga\pb}-
\epsilon^{\pb\pga} \Hhh^{\gga\pa}\,,
\ee
where $\Hhh^{\ga\pga}$ are the frame three-forms, we obtain
\bee \nn
  {H}^{\ga \gb}h^{\gga\pga}D^{L}{}_{\gga\pga}{C }_{\ga \gb} =2{\Hhh}^{\gb \pga}
  D^{L}{}^{\ga}{}_{\pga}{C }_{\ga \gb}\q
  \overline{ H}^{\pa \pb}h^{\gga\pga}D^{L}{}_{\gga\pga}{C }_{\pa \pb}
  =-2{\Hhh}^{\ga \pb}
  D^{L}{}_{\ga}{}^{\pa}{C }_{\pa \pb}.
\eee Hence  (\ref{Maxsum}) yields
  \bee \label{Max2}
D^{L}
\left(
\bar\eta {H}^{\ga \gb}{C}_{\ga \gb}(K|x)*  k-\eta\bar{H}^{\pa \pb}
 {C}_{\pa \pb} (K|x)*\bar k\right)\qquad\qquad\\ \nn= \bar\eta \eta  {\Hhh}^{\gga \pa}
 (
      2\bp_1{}_\pa \p_2{}_{\gga}- \bp_2{}_\pa  \p_2{}_{\gga}
      - \bp_1{}_\pa  \p_1{}_{\gga}+2\bp_2{}_\pa   \p_1{}_{\gga})
 \PPP_{\bf 0}
    (  y^1 ,  y^2 ; \bar  y^1   , \bar  y^2  ;K|x)  \,\big|_{ y^j=\by^j=0}  \eee
just reproducing the  Maxwell equations with a nonzero current.

 Substitution of   bilinear
  $\PPP_{\bf 0}\,$    \eq{Csumkbark'}  that by virtue of  inequality
   \eqref{sss} is     built from scalars or spinors
   gives
  \bee \label{Max2CC}
\sum_{j=0,1}D^{L}
\left(
\bar\eta {H}^{\ga \gb} C^{j,1-j}{}_{\ga \gb}(x)k^{1 +j}\bar k{}^{1-j}
-\eta\bar{H}^{\pa \pb}
   C^{j,1-j}{}_{\pa \pb} ( x)k^{ j}\bar k{}^{2-j}\right)\qquad\qquad
 \\ \nn=  \bar\eta \eta \sum_{j,\,l=0,1}  {\Hhh}^{\gga \pa}
 \Big(      2 (-1)^{ j}  C^{j,1-j}{}{}_\pa ( x)
 C^{l,1-l}{}_{\gga} ( x)
-   C^{j,1-j}{} ( x)
  C^{l,1-l}{}{}_\pa {}_{\gga} ( x)\quad
\\ \nn
      -   C^{j,1-j}{}{}_\pa {}_{\gga} ( x)
  C^{l,1-l}{} ( x)
+2(-1)^{1-j}
 C^{j,1-j}{} {}_{\gga} ( x)
  C^{l,1-l}{}{}_\pa ( x) \Big)
   k^{l+k}\bar k{}^{2-l-k} \,.\quad\eee

    \subsection{Spin $3/2$}
\label{examps=1.5j}
Using decomposition (\ref{jdecom})  and Eqs.~\eqref{canonical ressub==}, we obtain  from Eq.~(\ref{W2hhloc0})
 along with  \eq{Dad}, \eq{tw} 
 \bee
 \label{level-01.5}
&&D^L \go_{0\,,1}(0,\by;K|x)+ h(\p, \by) \go_{1\,,0}(y,0;K|x)
    \qquad\qquad\\ \nn&=&\eta\f{i}{4}
\bar{H}^{\pa\pb}{\bp_\pa \bp_\pb}
    {C}(0\,,\by;K|x)*\bar k+  H^{\ga\gb}{\p_\ga \p_\gb}
   \III^{ \f{3}{2}} _{ \half}{} \LLL_{2,1}(y^j,\by^j;K|x)\big|_{ y^j=\by^j=0}
\,,\\&&
   \label{level+01.5}
D^L \go_{1\,,0}(y,0;K|x)+h(y, \bp)\go_{0\,,1}(0,\by;K|x)
 \qquad\qquad\\ \nn&=&\bar\eta\f{i}{4}
H^{\ga\gb}{\p_\ga \p_\gb} C(y\,,0;K|x)*k
+ \bar{ H}^{\pa\pb}{\bp_\pa \bp_\pb}
         \III^{ \f{3}{2}} _{-\half}{} \LLL_{1,2}(y^j,\by^j;K|x)\big|_{ y^j=\by^j=0}\,,
\quad\eee
where     \bee\label{sp32}
  \III^{\f{3}{2}}_{-\half}{}&=&\half
 \int      d^2\gt
  \theta(\gt_1)\theta(\gt_2)   \gd'(1\!-\!\gt_1\!-\!\gt_2)  \big( \gt_1\NNN_1{}- \gt_2\NNN_2{}
  \big)
    \big(\gt_2\overline{\NNN}_1{}- \gt_1\overline{\NNN}_2{}\big) ^2\,,
   \\ \nn\III^{\f{3}{2}} _{ \half} &=&\half
 \int      d^2\gt
  \theta(\gt_1)\theta(\gt_2)   \gd'(1\!-\!\gt_1\!-\!\gt_2)  \big( \gt_1\NNN_1{}- \gt_2\NNN_2{}
  \big)^2
  \big(\gt_2\overline{\NNN}_1{}- \gt_1\overline{\NNN}_2{}\big)\q\\
&&\qquad \label{defN}
\NNN_j=y^\ga \p_j{}_\ga \q\qquad
\overline{\NNN}_j=\bar{y}^\pa \bp_j{}_\pa
 \qquad  \qquad
   \eee
 and, according to \eq {sss}, \eq{difJn} and  \eqref{canonical ressub==},
\bee\label{sp32J}
\LLL_{2,1} =  \frac{i}{8}\eta\bar\eta
 \Big(        \PPP_{\bf \half }  \,
       +\f{
          i\p_1{}_\gga\p_2{}^\gga}{2}
       \PPP_{\bf \f{3}{2}}\Big) \,
        \q \LLL_{1,2 } =  \frac{i}{8}
 \Big(        \PPP_{\bf -\half }  \,
       +\f{ i\bp_1{}_\pga\bp_2{}^\pga  }{2}
       \PPP_{\bf -\f{3}{2}}\Big) \,.
  \eee
 Representing one-forms $\go_{j\,,k}$   as
 \be\label{razdetom}\go_{j\,,k}{}=h^\ga{}^{\pb}\go{}_\ga{}_{\pb}{}_{j\,,k}
 \ee
 Eqs.~(\ref{level-01.5}), (\ref{level+01.5}), \eqref{sp32} yield spin-$ 3/2$
 massless equations in $AdS_4$ in the form
\bee
 \label{RarSh21}
     &&\Big (\bp_\pga D^L{}_{\ga\pb} \go_{0\,,1}{}_\ga{}^{\pb}(0,\by;K|x)
 -   \p_\ga \go_{1\,,0}{}_\ga{}_{\pga}(y,0;K|x) \Big )\,\big|_{ y=\by=0}
 \qquad \qquad\\ \nn&& =
   \frac{i}{8}\eta\bar\eta
  \Big(\f{1}{3}\p_1{}\p_1{}\bp_1{}
     + \p_2{}\p_2{}\bp_1{}
     - \p_1{}\p_1{}\bp_2{}
     -\f{1}{3}\p_2{}\p_2{}\bp_2{}
     -\f{2}{3}\p_1{}\p_2{}\bp_1{}
     +\f{2}{3}\p_1{}\p_2{}\bp_2{}
      \Big)_{\ga\ga;\pga}
\\ \nn&& \Big(        \PPP_{\bf \half }(y^j,\by^j;K|x)  \,
       +\f{  i\p_1{}_\gga\p_2{}^\gga}{2}
       \PPP_{\bf \f{3}{2}}(y^j,\by^j;K|x)\Big) \,\,\big|_{ y^j=\by^j=0}
         \, \quad\eee and complex conjugated.

         Using  inequality \eqref{sss},  substitution of   bilinear
  $\PPP  $   \eq{Csumkbark'} gives
    the Rarita-Schwinger equation  with   supercurrents on the right-hand side
   \bee && \label{RarSh21C}
     \Big ( \bp_\pb D^L{}_{\ga\pb} \go_{0\,,1}{}_\ga{}^{\pb}(0,\by ;K|x)
 -   \p_\ga \go_{1\,,0}{}_\ga{}_{\pb}(y,0 ;K|x)  \Big )\,\big|_{ y=\by=0}
 \\ &=&
   \frac{i}{4}\eta\bar\eta\sum_{j, l=0,1}  \nn\Big\{
 C^{j,1-j}{}_{ \pb}( x)
 C^{l,1-l}{}_{\ga\ga }( x)        +\f{          i(-1)^{ j}}{2}{
        C^{j,1-j}{}_{ \pb}{}_\gga( x)
 C^{l,1-l}{}_{\ga\ga }{}^\gga( x)}
       \\ && \nn   +
     (-1)^{  j}C^{j,1-j}{}{}_{\ga\ga}( x)
 C^{l,1-l}{}_{ \pb}( x){}        + \f{ i}{2}{
        C^{j,1-j}{}_{\ga\ga }{}_\gga( x)
 C^{l,1-l}{}_{ \pb}{}^\gga ( x)}
      \\ && \nn
       +  \f{1}{3}
  C^{j,1-j}{}_{\ga\ga;\pb}( x)
 C^{l,1-l}{}( x)        +\f{i(-1)^{j}}{6}{
        C^{j,1-j}{}{}_{\ga\ga;\pb}{}_\gga( x)
 C^{l,1-l}{}^\gga( x)}
       \\ && \nn
+\f{1}{3}
      C^{j,1-j}{}( x)
 C^{l,1-l}{}_{\ga\ga;\pb}{}( x)          + \f{ i} {6}{
        C^{j,1-j}{}_\gga( x)
 C^{l,1-l}{}_{\ga\ga;\pb}{}^\gga( x)}
       \\ && \nn
     -\f{2(-1)^{j}}{3}
     C^{j,1-j}{}{}_{\ga ;\pb}( x)
 C^{l,1-l}{}_{\ga }{}( x)        -\f{          i}{3}{
        C^{j,1-j}{}_{\ga ;\pb}{}_\gga( x)
 C^{l,1-l}{}_{\ga }{}^\gga ( x)}
       \\ && \nn-\f{2}{3}
     C^{j,1-j}{}( x){}_{\ga }
 C^{l,1-l}{}( x){}_{ \ga; \pb}        +\f{          i( -1 )^{1-j}}{3}{
        C^{j,1-j}{}_\gga{}_{\ga} ( x)
 C^{l,1-l}{}^\gga{}_{\ga ;\pb} } ( x)     \Big\}  k^{l+j}\bar k{}^{ -l-j}
       \, \quad
       \eee and complex conjugated.
   \subsection{Spin two}
\label{examps=20j}
   From Eq.~(\ref{W2hhloc0}), we    obtain by virtue of \eq{canonical ressub==}
\bee\nn
  \! D^{ad}\go(y,{\bar{y}};K|x)\!\!\!\! &=&\!\!\!\!
\left(      {H}^{\ga\gb}
 {\p_{\ga} \p_{\gb}}
 \III_0^{ 2 }\LLL_{ 2,2}(y^j,\by^j;K|x)
+   \bar{H}^{\pa\pb}
 {\bp_{\pa} \bp_{\pb}}
 \III_0^{ 2 }\LLL_{ 2,2}(y^j,\by^j;K|x) \right) \Big|_{y^j=\by^j=0}
\\ \label{CON1news2} &+&\f{ i}{4} \Big ( \eta \bar{H}^{\dga\pb}\bp_{\dga} \bp_{\dgb}\
{\bar C }(0,\overline{y};K| x) +\bar \eta H^{\ga\gb} \p_ {\ga} \p_{\gb}\
{C }(y,0;K| x)\Big )
\,,\eee where
\be \label{deftoksint2}
   \LLL_{2,2} =
\frac{i}{8}\eta\bar\eta \sum_{ 0\le \hf\le 2}
 \f{  1}{(1+  \hf   )!}
 \left[(i\p_1{}_\gga\p_2{}^\gga )^{\hf}\PPP_{  \hf }  +
        (i\bp_1{}_\pga\bp_2{}^\pga )^{ \hf   }     \PPP_{  -\hf}
    \right] \q   \ee
 \bee
   \label{seriescc0int2} 
  \III_0^2=
  \f{1}{  4!}
   \sum_{k=0}^2   \sum_{m=0}^2
  \f{(m+k)!(4 -m-k)!}{(2 -k)!k!(2 -m)!m!}  \big(  \NNN_1{}\big)^m
\big( -  \NNN_2{}\big)^{2 -m}
\big(-  \overline{\NNN}_2{}\big)^k
\big(   \overline{\NNN}_1{}\big)^{2 -k}  \,.
\quad
\eee
 In terms of decomposition   (\ref{jdecom}), this gives  in particular
\bee \nn
 D^L \go^{0\,,2}(0,\by;K |x)&=& - h^{\ga\pb} \by_\pb \p_\ga  \go^{1 ,1}(y,\by;K|x)
 + H^{\ga\gb}{\p_\ga \p_\gb}
   \III_0^2 \LLL_{2,2}(y^j,\by^j;K|x)\big|_{ y^j=\by^j=0}
\qquad
\\\ \label{level-1j}&+&
  \eta\f{i}{4}\bar{H}^{\pa\pb} {\bp_\pa \bp_\pb  }  C(0\,,\by\,;K|x)*\bar k
 \,,\quad\\
   \nn
D^L \go^{2\,,0}(y,0;K|x)
&=&  -h^{\ga\pb}  y_\ga \bp_\pb  \go^{1\,,1}(y,\by;K|x)\,
+   \bar{H}^{\pa\pb} {\bp_\pa \bp_\pb  }  \III_0^2 \LLL_{2,2}(y^j,\by^j;K|x)\big|_{ y^j=\by^j=0}
 \qquad\\\label{level+1j}&+&
    \bar\eta\f{i}{4}
H^{\ga\gb}{\p_\ga \p_\gb} C(y\,,0;K|x)*k
 .\quad\eee
 Using \eq{razdetom},
   this
    yields
\bee\!\!\!
 \label{spin2ur3j}
  \!D^L{}_{\gb\pb} \go^{0\,,2}{}_{\gb}{}^{\pb}
  &=&  {\p_\gb \p_\gb}
   \,\III_0^2 \LLL_{2,2} \big|_{ y^j=\by^j=0}
\!
+    \by_\pb \p_\gb
 \go^{1\,,1}{}_{\gb}{}^{\pb}
   \,,\quad\\
   \label{spin2ur4j}\!\!\!
\!D^L{}_{\gb\pb} \go^{2\,,0}{}^{\gb}{}_{\pb} 
&=&{\bp_\pa \bp_\pb  }     \,  \III_0^2 \LLL_{2,2} \big|_{ y^j=\by^j=0}
\!+    y_\gb
 {  \bp_\pb} \go^{1\,,1}{}^{\gb}{}_{\pb}
\,
\quad
\eee
giving
    the linearized Einstein equations
 \be
 \label{spin2ur3jj}
 \bp_\pa \bp_\pa
D^L{}_{\gb\pb} \go^{0,2}{}_{\gb}{}^{\pb}(0,\by ;K|x)
- 2    \p_\gb    \bp_\pa
 \go^{1 ,1}{}_{\gb}{}_{\pa}(y,\by;K|x)
  =  {\p_\gb \p_\gb}   \bp_\pa \bp_\pa
   \III_0^2 \LLL_{2,2}(y^j,\by^j;K|x)\big|_{ y^j=\by^j=0} \, \ee
accounting the contribution of the stress tensor.

From \eq{seriescc0int2} it follows
\bee\label{einst0}
 \bp_\pa\bp_\pa
D^L{}_{\gb\pb} \go^{0,2}{}_{\gb}{}^{\pb}(0,\by;K|x)
- 2    \bp_\pa
 \p_\gb
 \go^{1 ,1}{}_{\gb}{}_{\pa}(y,\by;K|x)
  =\qquad\qquad\\ \nn =
 \Big(  \p_1{}\p_1{}\overline{\p}_2{}\overline{\p}_2{}
+     \p_2{}\p_2{}\overline{\p}_1{}\overline{\p}_1{}
 -   \f{1}{2}\p_1{}\p_1{}\overline{\p}_1{}\overline{\p}_2{}
 +    \f{1}{6}\p_1{}\p_1{}\overline{\p}_1{}\overline{\p}_1{}
 - \f{1}{2} \p_1{}\p_2{}\overline{\p}_2{}\overline{\p}_2{}
+\f{2}{3}\p_1{}\p_2{}\overline{\p}_1{}\overline{\p}_2{}
\qquad\\ \nn
 -  \f{1}{2}\p_1{}\p_2{}\overline{\p}_1{}\overline{\p}_1{}
  +\f{1}{6}\p_2{}\p_2{}\overline{\p}_2{}\overline{\p}_2{}
 - \f{1}{2}\p_2{}\p_2{}\overline{\p}_1{}\overline{\p}_2{}
   \Big)_{\gb\gb\pa\pa}  \,   \LLL_{2,2}(y^j,\by^j;K|x)\big|_{ y^j=\by^j=0}\,.\qquad 
    \eee
Substitution  of $\PPP$ \eq{Csumkbark'}  into  $\LLL_{2,2}$
yields \bee\nn
   \LLL_{2,2} =      \frac{i}{8}\eta\bar\eta\sum_{j , l=0,1}
       \Big\{    C^{j,1-j}{}
                C^{l,1-l}{}
 +\f{ i(-1)^{j}
                C^{j,1-j}{}_\gga
                C^{l,1-l}{}^\gga }{2}
 -\f{
                C^{j,1-j}{}_\gga{}_\gd
                C^{l,1-l}{}^\gga{}^\gd }{3!}\qquad
       \\ \label{tokspin2}
         +\f{ i(-1)^{j}
                C^{j,1-j}{}_\pga
                C^{l,1-l}{}^\pga }{2}
 -\f{
                C^{j,1-j}{}_\pga{}_\pd
                C^{l,1-l}{}^\pga{}^\pd }{3!} \Big\}(x)    k^{l+j}\bar k{}^{ -l-j}.
 \quad      \eee
 Hence from (\ref{deftoksint2}),  \eq{einst0}
and \eq{tokspin2}
it  follows
 \bee\label{einstexam}
\bp_\pa\bp_\pa
D^L{}_{\gb\pb} \go^{0,2}{}_{\gb}{}^{\pb}(0,\by;K|x)
- 2    \bp_\pa
 \p_\gb
 \go^{1 ,1}{}_{\gb}{}_{\pa}(y,\by;K|x)
  =\qquad\qquad\\ \nn =   \frac{i}{8}\eta\bar\eta\sum_{  j,l=0,1}
  \Big\{
    C^{j,1-j}{}_{\gb\gb } C^{l,1-l}{}_{ \pa\pa}
    +
 C^{j,1-j}{}_{ \pa\pa} C^{l,1-l}{}_{\gb\gb }
 - \half (-1)^{1-j} 
  C^{j,1-j}{}_{\gb\gb\pa } C^{l,1-l}{}_{ \pa}\\ \nn
 +    \f{1}{6}
  C^{j,1-j}{}_{\gb\gb\pa\pa} C^{l,1-l}{}
 - \half
   (-1)^{ j}C^{j,1-j}{}_{\gb } C^{l,1-l}{}_{\gb \pa\pa}
 - \f{2}{3}
 C^{j,1-j}{}_{\gb \pa}   C^{l,1-l}{}_{\gb \pa}
  +\f{1}{6}
   C^{j,1-j}{} C^{l,1-l}{}_{\gb\gb\pa\pa}
\\ \nn\\ \nn - \f{1}{2}(-1)^{ j}
  C^{j,1-j}{}_{ \gb\pa\pa}   C^{l,1-l}{}_{\gb }
 -\f{1}{2}(-1)^{1- j}
   C^{j,1-j}{}_{ \pa}   C^{l,1-l}{}_{\gb\gb\pa }
  \Big\}(x) k^{l+j}\bar k{}^{2-l-j}  \,   \quad+ \ldots
    \eee
  with  the stress tensor of  massless fields of spins $0,\,\,1/2$
and $1$. Ellipses denotes  other currents, that depend on   massless fields of spins  $s\le2$ and respect   \eq{sss}.

 \subsection{  Higher  spins }
\label{examps=s0j}

  \subsubsection{Integer spins}

  Using the
decomposition (\ref{jdecom}) for $\go$ and Eqs.~\eqref{canonical ressub==}, it follows
 from (\ref{W2hhloc0})
 that
   \eq{Dad}
 yields
  \bee\label{levels0}
D^L \go_{\pss-1\,,\pss-1}(y,\by;K|x)
=  h(\p,\by) \go_{\pss \,,\pss -2}(y,\by;K|x) \!+
 h(y,\bp)   \go_{\pss -2\,,\pss }(y,\by;K|x)
 \,,\qquad\qquad
\\\nn\\ \label{level+1s}
D^L \go_{\pss \,,\pss -2}(y,\by;K|x)=- 
h(y,\bp) \go_{\pss-1\,,\pss-1}(y,\by;K|x)
   \qquad\qquad\qquad\qquad \qquad \\ \nn
  -h(\p,\by)\go_{\pss +1\,,\pss -3}(y,\by;K|x)
+   \bar{H}^{\pa\pb}
 {\p_{\pa} \p_{\pb}}
 \III_0^{\pss }\LLL_{\pss,\pss} (y^j,\by^j;K|x)\big|_{ y^j=\by^j=0}
\q\\ \nn\\
 \label{level-1s}\!
D^L \go_{\pss -2\,,\pss }(y,\by;K|x)=-h(\p,\by)  \go_{\pss-1\,,\pss-1}(y,\by;K|x)
  \qquad\qquad\qquad\qquad\qquad \\\nn-   h(y,\bp)\go_{s-3\,,s+1}(y,\by;K|x)\!+
  H^{\ga\gb}
 {\p_{\ga} \p_{\gb}} \III_0^{\pss }\LLL_{\pss,\pss} (y^j,\by^j;K|x)\big|_{ y^j=\by^j=0}
 \q \eee
   \bee \label{deftoksint}
    \LLL_{\pss,\pss} &=&
\frac{i}{8}\eta\bar\eta\left\{ \sum_{ 0\le \hf\le \pss}\Big[
 \f{( i\p_1{}_\gga\p_2{}^\gga )^{ |{\hf}| }}{(\pss+ |\hf| -1)!}
                \PPP_{\hf }\Big] \,
+    \sum_{ 0< -\hf\le \pss} \Big[
 \f{( i\bp_1{}_\pga\bp_2{}^\pga )^{ |{\hf}|  } }{(\pss+ |\hf| -1)!} \PPP_{\hf}\Big]
    \,\right\} \q   \\
  \label{seriescc0}
  \III_0^\pss&=&
 \int      d^2\gt
  \theta(\gt_1)\theta(\gt_2)   \gd'(1\!-\!\gt_1\!-\!\gt_2) \f{\big( \gt_1\NNN_1{}- \gt_2\NNN_2{}
  \big)^{\pss  }}{ \pss !}    \frac{
  \big(\gt_2\overline{\NNN}_1{}- \gt_1\overline{\NNN}_2{}\big) ^{\pss  } }
 {(\pss-1) \pss  }\,.
 \qquad\eee
{}From here it follows that 
\bee\label{Levs0}
h^{\gga\pga}h^{\ga\pa}D^L_{\gga\pga} \go_{\pss-1\,,\pss-1}{}_{\ga\pa}
=
-h(\p,\by)h^{\ga\pa}  \go_{\pss \,,\pss -2}{}_{\ga\pa}  \!-
h(y,\bp)h^{\ga\pa}   \go_{\pss -2\,,\pss }{}_{\ga\pa}
\,,
\eee
\bee \label{Lev+1s}
h^{\gga\pga}h^{\ga\pa}D^L_{\gga\pga}  \go_{\pss \,,\pss -2}{}_{\ga\pa} =-
h(y,\bp)h^{\ga\pa}  \go_{\pss-1\,,\pss-1}{}_{\ga\pa}
 + \qquad\qquad\qquad\qquad \qquad\\ \nn
 - h(\p,\by)h^{\ga\pa}  \go_{\pss +1\,,\pss -3}{}_{\ga\pa}
+     \bar{H}^{\pa\pb}
 {\p_{\pa} \p_{\pb}}
 \III_0^{\pss }\LLL_{\pss,\pss}  \big|_{ y^j=\by^j=0}
,\eee
\bee
 \label{Lev-1s}\!
h^{\gga\pga}h^{\ga\pa}D^L_{\gga\pga}   \go_{\pss -2\,,\pss }{}_{\ga\pa}
 =- h(\p,\by)   \go_{\pss-1\,,\pss-1}
 +\qquad\qquad\qquad\qquad\qquad \\\nn
 - h(y,\bp)h^{\ga\pa}  \go_{\pss -3\,,\pss +1}{}_{\ga\pa} \!
 +  H^{\ga\gb}
 {\p_{\ga} \p_{\gb}} \III_0^{\pss }\LLL_{\pss,\pss}   \big|_{ y^j=\by^j=0}
 .
\quad\eee
Hence
  \bee%
 \label{Luv-1s1}\!
 D^L_{\ga\pga}   \go_{\pss -2\,,\pss }{}{}_{\ga}{}^{\pga}
=-   \by_\pb \p_\ga  \go_{\pss-1\,,\pss-1}{}_{\ga}{}^{\pb}
  -  {y}_\ga \bp_\pb    \go_{\pss -3\,,\pss +1}{}_{\ga}{}^{\pb}+
  {\p_{\ga} \p_{\ga}} \III_0^{\pss }\LLL_{\pss,\pss} \big|_{ y^j=\by^j=0}
 \, ,
\qquad\\\label{Luv-1s2}
 D^L_{\gga\pb}  \go_{\pss \,,\pss -2}{}^{\gga}{}_{\pb} =-
 {y}_\ga \bp_\pb    \go_{\pss-1\,,\pss-1}{}^{\ga}{}_{\pb}
  -   \by_\pb \p_\ga  \go_{\pss +1\,,\pss -3}{}^{\ga}{}_{\pb}
 +
 {\p_{\pa} \p_{\pb}}
 \III_0^{\pss }\LLL_{\pss,\pss}  \big|_{ y^j=\by^j=0}
\,.
\qquad\eee
Integrating over $\gt$ in \eq{seriescc0}
and  substituting      $\PPP$ \eq{Csumkbark'}  into $ \LLL_{\pss,\pss} $  \eq{deftoksint}
      one obtains
 \bee \label{LJns}
   \III_0^{\pss }   \LLL_{\pss,\pss}  =
i\eta\bar\eta
   \f{(\pss-2)!}{8 (2\pss)!}\!
   \sum_{k,m\in[0,\pss] } \! \!
     \f{(m+k)!(2\pss-m-k)!}{(\pss-k)!k!(\pss-m)!m!}  \big(  \NNN_1{}\big)^m
\big(\!-\!  \NNN_2{}\big)^{\pss-m}
\big(\!-\!  \overline{\NNN}_2{}\big)^k
\big(   \overline{\NNN}_1{}\big)^{\pss-k}\qquad\\ \nn
\Big\{\sum_{ 0\le  \hf\le \pss}
\f{1}{(\pss+  \hf  -1)!}( i\p_1{}_\gga\p_2{}^\gga )^{  \hf   }
 \sum_{j,l=0,1}
C^{j,1-j}(Y^1|x) k^j\bar k{}^{1-j}
  C^{l,1-l}(Y^2|x) k^l\bar k{}^{1-l}
 \rule{65pt}{0pt}\\ \nn\!\!\!\!\!\!+
   \sum_{ 0<  \hf\le \pss}
\f{1}{(\pss+  \hf  -1)!}( i\bp_1{}_\pga\bp_2{}^\pga )^{  {\hf}   }
 \sum_{j,l=0,1}
C^{j,1-j}(Y^1|x) k^j\bar k{}^{1-j}
  C^{l,1-l}(Y^2|x) k^l\bar k{}^{1-l}
 \Big\}         \Big|_{ Y^j=0} \,.
\qquad
\eee
Substitution  of  
   $\III^\pss_{0 }{}  \LLL_{\pss ,\pss } $   \eq{LJns} into \eq{Luv-1s1} and \eq{Luv-1s2}
   yields  integer-spin field equations  in $AdS_4$ with the conformal currents.

To obtain the dynamical spin-$\pss$ equations with the current corrections
it remains to project away the terms, that contain $\omega^{\pss -3,\pss +1}$ and
$\omega^{\pss +1,\pss -3}$. This is achieved by the contraction of free indices
with $y^\ga y^\ga$  in
(\ref{Luv-1s1}) and    $\bar{y}^\pb \bar{y}^\pb$  in
(\ref{Luv-1s2}).
The resulting equations describe the contribution of HS currents
to the right-hand-sides of Fronsdal's equations
in $AdS_4$.

Let the constituent fields in $J$ have helicities $h_1 $ and  $h_2$.
Since each helicity-$h$ field $C$  in flat limit contains $|h|$
space-time derivatives of $\go_{s-1,s-1}$ due to Central-on-shell theorem
\eq{CON1k},
then from \eq{fltw} and \eq{Luv-1s1}-\eq{LJns} it follows that
the number of space-time derivatives in the respective flat limit
vertices
is  $\pss+|h_1+h_2|$ for any helicities obeying \eq{sss}, \ie
$\pss\ge |h_1|+|h_2|$,
 just reproducing  the results of Metsaev \cite{Metsaev:2005ar}.

  \subsubsection{Half-integer spins}
In terms of decomposition (\ref{jdecom}),
for any   half--integer $\pss> 1$, it follows
 from Eqs.~\eqref{canonical ressub==}, (\ref{W2hhloc0}) that
 \bee
  \label{level-0p.5}
  &&D^L \go_{{[s]-1}\,,{[s]}}(y,\by;K|x)=
- h(y ,\bp)  \go_{{[s]}-2\,,{[s]}+1}(y,\by;K|x)
 \\ \nn
\!\!\!&-&\! h(\by,\p)  \go_{{[s]}\,,{[s]-1}}(y,\by;K|x)
 +
 H^{\ga\gb}{\p_\ga \p_\gb} \III^\pss{}_{ \half}{} \LLL_{\pss+\half,\pss-\half}
  (y^j,\by^j;K|x)\big|_{ y^j=\by^j=0}
  \,\eee and complex conjugated,
where       \be \label{halfpl}
 \LLL_{\pss+\half,\pss-\half} =  \frac{i}{8} \sum_{ \half\le \hf\le \pss}
 \f{1}{(\pss+ |\hf| -1)!}
  \Big[       (   i\p_1{}_\gga\p_2{}^\gga)^{[|{\hf}|]}
       \PPP_{\hf }\Big]   \q
  \ee
\be \label{Lambdahalfp}
  \III^\pss{}_{ \half} =
 \int      d^2\gt
  \theta(\gt_1)\theta(\gt_2)   \gd'(1\!-\!\gt_1\!-\!\gt_2) \f{\big( \gt_1\NNN_1{}- \gt_2\NNN_2{}
  \big)^{\pss +\half }}{({\pss-\half })(\pss +\half)}    \frac{
  \big(\gt_2\overline{\NNN}_1{}- \gt_1\overline{\NNN}_2{}\big) ^{\pss -\half } }
 {(\pss-\half)!  }\,
  .\qquad \ee
 Hence
\be
  \label{Lovel-0p.5}
   D^L_{\ga\pga} \go_{{[s]-1}\,,{[s]}}{}_{\ga}{}^{\pga}  =-  {y}_\ga \bp_\pb
 \go_{{[s]}-2\,,{[s]}+1} {}_{\ga}{}^{\pb}
-    \by_\pb \p_\ga             \go_{{[s]}\,,{[s]-1}}{}_{\ga}{}^{\pb}
  +
  {\p_\ga \p_\ga} \III^\pss{}_{ \half}{} \LLL_{\pss+\half,\pss-\half}   \big|_{ y^j=\by^j=0}
  \,.\qquad\ee
  Integrating over $\gt$ in  \eq{Lambdahalfp}
and  substituting     $\PPP$ \eq{Csumkbark'}  into $ \LLL_{\pss,\pss} $
     yields by virtue of  \eq{halfpl}, \eq{Lambdahalfp}
 \bee \label{Lambdahalfintp} 
  \III^\pss{}_{ \half}\LLL_{\pss+\half,\pss-\half} (y^j,\by^j;K|x)\big|_{ y^j=\by^j=0} =  {i}  \eta\bar\eta
   \f{(\pss-3/2)!}{8(2\pss)!}
   \sum_{   \half<  \hf\le \pss} \f{1}{(\pss+  \hf  -1)!}
 \qquad\\ \nn
\sum_{k,m=0}^\pss  \f{(m+k)!(2\pss-m-k)!}{ (\pss+\half-k)!k!({\pss-\half-m})!m!}
 \big(  \NNN_1{}\big)^k
\big( - \NNN_2{}\big)^{\pss+\half-k}
\big( -\overline{\NNN}_2{}\big)^m
\big(   \overline{\NNN}_1{}\big)^{\pss-\half-m}
\qquad\\ \nn        (   i\p_1{}_\gga\p_2{}^\gga)^{  \hf-\half }
 \sum_{j,l=0,1}
C^{j,1-j}(Y^1|x) k^j\bar k{}^{1-j}
  C^{l,1-l}(Y^2|x) k^l\bar k{}^{1-l}
\,\Big|_{ Y^j =0}   \,.    \qquad\eee

Substitution of
   $\III^\pss{}_{ \half}{}  \LLL_{\pss+\half,\pss-\half} $   \eq{Lambdahalfintp} into \eq{Lovel-0p.5}
gives the  half-integer spin   equations  in $AdS_4$
with the conformal currents. Complex conjugated equations are analogous.

Projecting away the terms,   with the extra fields $\go_{{[s]}-2\,,{[s]}+1}$
and $\go_{{[s]}+1\,,{[s]}-21}$ by   contracting   free indices
with $y^\ga y^\ga$ and  $\bar{y}^\pb \bar{y}^\pb$, respectively,
gives the Fang-Fronsdal field equations \cite{Frfhs} in $AdS_4$
with the conformal currents on the right-hand-sides.

\section{Conclusion}
\label{conc}
We have derived  current sources to the right-hand side of  field equations on
massless fields of all spins resulting from the nonlinear field HS equations
of \cite{more}. Our results extend those obtained by of one of us
\cite{Vasiliev:2016xui} for current interactions in the    zero-form sector.
The derivation agrees with that of \cite{Vasiliev:2016xui}
in many respects.

First of all, in agreement with the conclusion of \cite{Vasiliev:2016xui},
the bilinear (current) corrections turn out to be independent of the phase of the parameter
$\eta$ in the HS theory, depending only on $\eta \bar\eta$. Naively this
contradicts the standard expectation that the HS theory with different phases
of $\eta$ correspond to different boundary conformal theories with Chern-Simons
fields. However, as explained in  \cite{Vasiliev:2016xui,Didenko:2017lsn}, the proper
dependence on the phase of $\eta$ in the HS $AdS/CFT$ correspondence results from
 the phase-dependence of the linear terms in the HS equations
 upon transition to the genuine Weyl tensors.
For general $\eta$ our vertex contains both parity even and
parity odd parts, which appear in HS models with general $\eta$.

We not only reduced the HS current interactions to the local form with finite number
of derivatives for any three spins, but also found  its canonical form with the
minimal number of derivatives and zero HS torsion. The resulting coupling constants
are nonzero being uniquely determined in terms of the single
HS coupling constant $\eta\bar\eta$.

 The detailed computation
of the resulting boundary correlators is presented in
\cite{Didenko:2017lsn} based on the zero-form results of \cite{Vasiliev:2016xui}
 (for a special case see also \cite{Sezgin:2017jgm}). The results of this paper allow
 one to extend this analysis to the one-form (\ie gauge field) sector checking in
 particular  whether the nonlinear deformation of this paper
 matches the cubic vertex derived in \cite{Sleight:2016dba} from the holographic analysis  for the $A$-model.
 This problem was considered in \cite{misuna} where it  shown
 that the coefficients in the vertices resulting from our analysis match those of
 \cite{Metsaev:1991mt,Sleight:2016dba}.  Let us stress that the current interactions
 derived in this paper extend the vertex of \cite{Fradkin:1987ks,Fradkin:1986qy,Sleight:2016dba} to the parity non-invariant
 vertices holographically dual to the HS theory with an arbitrary phase parameter $\eta$
 (see \cite{Didenko:2017lsn,misuna}).

The conclusion  that the contribution to the currents
proportional to $\eta^2$ should vanish fits the conjecture of \cite{more} that
the HS theory with ($\eta=0$)$\bar \eta=0$ is the \mbox{(anti-)}self-dual  HS gauge
theory. In that case it should describe the zero-form curvatures with
only positive or only negative helicities which cannot contribute to nontrivial
currents for the same reason why the amplitudes with helicities of the
same sign cannot be nonzero. Hence, our results confirm the conjecture that
the HS theory at ($\eta=0$)$\bar \eta=0$ is (anti)self-dual.

The obtained results provide a basis for understanding the proper general
setup for the systematic derivation of minimally nonlocal perturbative
corrections to nonlinear HS equations. This issue is
considered in   \cite{1707.03735}.
The form of the results obtained in
\cite{Vasiliev:2016xui} and in this paper
 demonstrates in particular that this
prescription should allow a proper formulation in the geometric terms of polyhedra
associated with the integration parameters $\tau_i$.

 \section*{Acknowledgements}
We are grateful to Slava Didenko for critical comments on the manuscript,  Nikita
Misuna and Ruslan Metsaev for useful discussions, and Karapet Mkrtchyan for the correspondence.
 We would like to thank for hospitality the MIAPP programme "Higher Spin Theory and Duality"
in May 2016 and Sergey Kuzenko, the School of Physics at the University
of Western Australia, in November 2016  during the
various stages of the project, as well as
 the Galileo Galilei Institute for Theoretical Physics (GGI) for the hospitality and INFN for
 partial support during the completion of this work, within the program "New Developments in AdS3/CFT2 Holography".
 The work of MV is partially supported by the  ARC Discovery Project DP160103633 and
by a grant from the Simons Foundation. We acknowledge  a partial support from  the Russian Basic
Research Foundation Grant No 17-02-00546.

 \newcounter{appendix}
\setcounter{appendix}{1}
\renewcommand{\theequation}{\Alph{appendix}.\arabic{equation}}
\addtocounter{section}{1} \setcounter{equation}{0}
 \renewcommand{\thesection}{\Alph{appendix}.}
 \addtocounter{section}{1}
\addcontentsline{toc}{section}{\,\,\,\,\,\,\,Appendix A.  {Useful formulas}}
\section*{Appendix A.  Useful formulas}\label{Appendix 1}
In the analysis of HS perturbations it is convenient to use
   the following generalized beta-function formula:
 \bee\label{inttau1-tau01pp}
 \int\! d\gt^m \gd^{(k)} \!\left(\!1\!-\!\sum\limits_{i=1}^m\gt_i\!\right)
  \prod\limits_{i=1}^m\theta(\gt_i)\gt_i^{n_i}  =
 \f{ \prod\limits_{i=1}^m  {n_i}! }{ \Big(\!\sum\limits_{i=1}^m n_i +m \!-\!1\!-\!k\!\Big)!}
 \q\forall n_i, k\ge0 \,.\eee

Now we reproduce some of formulas of \cite{Didenko:2015cwv}  most relevant to
the analysis of this paper. Let
$  \phi^{AB}Y_{A}Y_{B} =4i\phi_{AdS}$
 ,  where $\phi_{AdS}$ is defined in
\eq{fads}.
Then
 \begin{eqnarray}\label{adj}
&&\triangle_{ad}^{*}\A\left(Z;Y;\theta\right)   \! = \!   \dfrac{i}{2}\int d\mu d\varphi
d\chi dUdV\intop_{0}^{1}\dfrac{d\gt}{\gt}
\exp\left\{ \chi_{A}\varphi^{A}+iU_{A}V^{A}\right\}   \\\nn&&
       \exp\left\{ \mu \gt\chi_{A}Z^{A}+\dfrac{i}{2}\left(1-\gt\right)\phi^{BC}\chi_{B}U_{C}\right\}
 \A\left(\gt Z;Y+V;\gt\theta +\varphi\right),
  \eee
  \be\label{Hadj}
\Hh_{ad}J\left(Z;Y;\theta\right)  \! = \!  \int d\varphi d\chi dUdV
\exp\left\{ \chi_{A}\varphi^{A}\!+\!iU_{B}V^{B}\!+\!\dfrac{i}{2}\phi^{BC}\chi_{B}U_{C}\right\}
 \A\left(0;Y\!+\!V;\varphi\right),
\qquad\ee
\bee\label{twadj}&&
\triangle_{tw}^{*}\A\left(Z;Y;\theta\right) \stt k  \!  = \!   \dfrac{i}{2}
\int d\mu d\sigma d\rho dUdVdPdQ
\intop_{0}^{1}\dfrac{d\gt}{\gt}
 \quad \\\nn&&
\exp\left\{ \mu \gt\sigma_{A}Z^{A}+\gs_{A}\gr^{A}+iU_{A}V^{A}+iP_{A}Q^{A}\right\}
 \quad \\&&
        \exp\left\{ -\dfrac{i}{2}\left(1-\gt\right)\omega^{AB}\sigma_{A}V_{B}+\dfrac{i}{2}\left(1-\gt\right)h^{AB}
 \sigma_{A}\left(Y_{B}+\dfrac{1}{2}U_{B}+\dfrac{1}{2}\left(1+\gt\right)Q_{B}\right)\right\}
  \nonumber \\\nn
     &&  \A\left(\gt Z+P;Y+U;\gt\theta+\rho\right)\stt k,
 \eee\bee
 \label{antidz}&&
\Hh_{tw}\A\left(Z;Y;\theta\right)\stt k   \! \! = \! \!  \int d\sigma d\rho dUdVdPdQ
\exp\left\{ \rho_{A}\sigma^{A}+iU_{A}V^{A}+iP_{A}Q^{A}\right\}   \\\nonumber
    &&  \exp\left\{ -\dfrac{i}{2}\omega^{AB}\sigma_{A}V_{B}+\dfrac{i}{2}h^{AB}
 \sigma_{A}\left(Y_{B}+\dfrac{1}{2}U_{B}+\dfrac{1}{2}Q_{B}\right)\right\} \A\left(P;Y+U;\rho\right)\stt k.\nn
\end{eqnarray}

\addtocounter{appendix}{1} \setcounter{equation}{0}
  \addtocounter{section}{1}

 \section*{Appendix B. Alternative redefinitions}

\addcontentsline{toc}{section}{\,\,\,\,\,\,\,Appendix B. Alternative redefinitions}

\label{Appendix 2}

Let
\be\label{sourceredifeta1}
  X (\PPP)  =
  \! \int d^3\bgt    d^3\gt \,h^{\ga\pb}X_{\ga\pb}(\gt,\bgt)
  \exp\big(  \gt_3\p_1{}_{\ga}\p_2{}^{\ga}+\bgt_3\bp_1{}_{\pa}\bp_2{}^{\pa}
  \big)  \PPP(   \gt_1y,  -\gt_2 y ,   \bgt_1\by,-\bgt_2\by;K|x) \,,\qquad
\ee
where
\bee\label{sourccoef}
X_{\ga\pb}=a y_{\ga}\by_{\pb}+
y_{\ga}\sum_i \bar b_i  \bp_i{}_{\pb}
+\sum_i
 b_i \p_i{}_{\ga}\by_{\pb}
+\sum_{i,j}  g_{i\,j} \p_i{}_{\ga}\bp_j{}_{\pb}\,.\eee
We will look for a  solution to
\be\label{eqdadX}
\D_{ad}X = \GGG_{\eta\bar\eta } + G^{loc},
\ee
where
$\GGG_{\eta\bar\eta }$ is given by (\ref{etabetashift'}), while $G^{loc}$ is some local vertex to be found.

For simlicity we set    $i\eta\bar\eta=-4$ in (\ref{etabetashift'}).
Denote
\bee\label{den2}&&A_j:=i \dt_j a+\dt_3\tilde b_j\q
B=\dt_1 b_1-\dt_2 b_2\q\tilde b_2=b_1\q
\tilde b_1=b_2\q\\ \nn
  \label{den3}&&
G_{kj}=\dt_3\tilde g_{kj}+i\dt_k \bar b_j \q F_j=\dt_1 g_{1j}-\dt_2 g_{2j}\q
 \tilde g_{2j}=  g_{1j}\q\tilde g_{1j}=  g_{2j}\q\\
 \label{den1}\nn&&\f{\p}{\p \gt_j}:=\dt_j\q \f{\p}{\p \bgt_j}:=\pt_j\,.\eee

By virtue of the  Fierz (Schouten) relations expressing the two-componentness of spinorial
indices in the form
\bee\label{Firtz0}
(i\by_\pa\f{\p}{\p \bgt_3}+\bp_1{}_\pa\f{\p}{\p \bgt_2}\!+\!\bp_2{}_\pa\f{\p}{\p \bgt_1})  \exp\big(  \gt_3\p_1{}_{\ga}\p_2{}^{\ga}+\bgt_3\bp_1{}_{\pa}\bp_2{}^{\pa}
  \big)  \PPP(   \gt_1y,  \!-\!\gt_2 y ,   \bgt_1\by,-\!\bgt_2\by;K|x)=0\,.\qquad
\eee
Then \be\nn\F \exp\big(  \gt_3\p_1{}_{\ga}\p_2{}^{\ga}+\bgt_3\bp_1{}_{\pa}\bp_2{}^{\pa}
  \big)  \PPP(   \gt_1y,  -\gt_2 y ,   \bgt_1\by,-\bgt_2\by;K|x)=0\,  \ee
  for any $\F$ of the form
\be\label{Firtz}
\F=\half
h_{\mu}{}^{\pa}h^{\mu\pb}  \Big( \ga \by_\pb+ \gb_1 \bp_1{}_\pb+ \gb_2 \bp_2{}_\pb\Big)
\Big(i\by_\pa\f{\p}{\p \bgt_3}+\bp_1{}_\pa\f{\p}{\p \bgt_2}+\bp_2{}_\pa\f{\p}{\p \bgt_1}\Big) \,.
 \ee
In this setup
\bee\label{DX+F}
\Big(D_{ad}X+\F\Big)\big|_{\bar H}=
-\half \bar H^{\pa\pb}
   \int \int d^3\bgt    d^3\gt \,
\rule{100pt}{0pt}\\ \nn %
\Big\{\Big[
B\by_\pa+F_j\bp_j{}_\pa\Big]\Big[(i\gt_1\bgt_1-i\gt_2\bgt_2)\by_\pb+
( \bgt_1+\gt_2\bgt_3)\bp_1{}_\pb-( \bgt_2+\gt_1\bgt_3)\bp_2{}_\pb
\Big]\qquad\\\nn
+i
\Big[
A_1\by_\pa+G_{1j}\bp_j{}_\pa\Big]\Big[( \gt_1+\gt_3\bgt_2)\by_\pb+i
(  \gt_3\bgt_3-1)\bp_1{}_\pb
\Big]\qquad\\\nn
+i
\Big[
A_2\by_\pa+G_{2j}\bp_j{}_\pa\Big]\Big[( \gt_2+ \gt_3\bgt_1)\by_\pb+i
(  \gt_3\bgt_3-1)\bp_2
{}_\pb
\Big]\qquad\\ \nn-\Big(i\by_\pa  \pt_3 +\bp_1{}_\pa {\pt_2}+\bp_2{}_\pa\pt_1\Big)
\Big( \ga \by_\pb+ \gb_1 \bp_1{}_\pb+ \gb_2 \bp_2{}_\pb\Big)
 \Big\}\,\\ \nn
 \exp\big(  \gt_3\p_1{}_{\ga}\p_2{}^{\ga}+\bgt_3\bp_1{}_{\pa}\bp_2{}^{\pa}
  \big)  \PPP(   \gt_1y,  -\gt_2 y ,   \bgt_1\by,-\bgt_2\by;K|x) \,.\qquad
 \eee
 We found two   solutions to this problem.
 The solution I, being technically more involved,  which uses  Fierz  relations in full generality,
 is simpler methodologically. The solution II, which is based on the results of
 \cite{Gelfond:2010pm},
 has simpler form but contains a  $\gd$-function of some nonlinear argument that demands a proper
 definition eventually leading to the  simple expression (\ref{sourceredifeta0}).

\subsection*{Solution I}
One can make sure that the following    coefficients solve  the problem:
(the factor of $\Upsilon
$ is implicit)
\bee\label{rescoef}
a&=&-\half\Big\{\gd(\gt_3)\big[( \bgt_1+ \gt_2\bgt_3)\gt_1\gd(\bgt_2)+( \bgt_2+ \gt_1\bgt_3)\gt_2\gd(\bgt_1) \big]
\\ \nn& &\quad\,\,\,+\gd(\bgt_3)\big[( \gt_1+ \gt_3\bgt_2)\bgt_1\gd(\gt_2)+( \gt_2+ \gt_3\bgt_1)\bgt_2\gd(\gt_1)\big]\Big\}
 \gd(\gs)\gd(\bar \gs)\q\\ \nn
b_1&=&\f{i}{2}( \gt_1+ \gt_3\bgt_2)\gd(\gt_2)\gd(\gs)\gd'(\bar \gs) \q
 b_2=\f{i}{2}( \gt_2+ \gt_3\bgt_1)\gd(\gt_1)\gd(\gs)\gd'(\bar \gs) \q
\\ \nn
\bar{b}_1&=&\f{i}{2}( \bgt_1+ \bgt_3\gt_2)\gd(\bgt_2)\gd'(\gs)\gd(\bar \gs) \q
 \bar{b}_2=\f{i}{2}( \bgt_2+ \bgt_3\gt_1)\gd(\bgt_1)\gd'(\gs)\gd(\bar \gs) \q
\eee\bee \nn
g_{12}&=& (- 1+ \gt_3\bgt_3)\gd(\gt_2)\gd(\bgt_1)\gd(\gs)\gd(\bar \gs) \q
 g_{21}= (- 1+ \gt_3\bgt_3)\gd(\gt_1)\gd(\bgt_2)\gd(\gs)\gd(\bar \gs) \q
\\ \nn
g_{11}&=&\half( 1- \gt_3\bgt_3)
\big[ \gd(\bgt_2)\gd'(\gs)\gd(\bar \gs)+\gd(\gt_2) \gd(\gs)\gd'(\bar \gs)
-\gd(\gt_2)\gd(\bgt_2)\gd(\gs)\gd(\bar \gs)\big]
\q
\\ \nn
g_{22}&=&\half( 1- \gt_3\bgt_3)
\big[ \gd(\bgt_1)\gd'(\gs)\gd(\bar \gs)+\gd(\gt_1) \gd(\gs)\gd'(\bar \gs)
-\gd(\gt_1)\gd(\bgt_1)\gd(\gs)\gd(\bar \gs)\big]
\q\eee
\bee\label{alfa-beta}
 \ga&=&\f{i}{2}\big[\bgt_1(\gt_1+\gt_3\bgt_2)\gd(\gt_2)
+\bgt_2(\gt_2+\gt_3\bgt_1)\gd(\gt_1)\big]\gd'(\gs)\gd(\bar \gs)\q
\\ \nn
 \gb_1&=& \half\big[\bgt_1(\bgt_1-\bgt_2)\gd(\gt_1)\gd(\gt_2)\gd(\gs)\gd(\bar \gs)
+( 1- \bgt_3\gt_3)\bgt_1 \gd(\gt_2)\gd'(\gs)\gd(\bar \gs)
\big]\q
\\ \nn
 \gb_2&=& \half\big[\bgt_2(\bgt_2-\bgt_1)\gd(\gt_1)\gd(\gt_2)\gd(\gs)\gd(\bar \gs)
+( 1- \bgt_3\gt_3)\bgt_2 \gd(\gt_1)\gd'(\gs)\gd(\bar \gs)
\big]\q
\eee
 \bee\label{bar-alfa-beta}
\bar\ga&=&
\f{i}{2}\big[\gt_1(\bgt_1+\bgt_3\gt_2)\gd(\bgt_2)
+\gt_2(\bgt_2+\bgt_3\gt_1)\gd(\bgt_1)\big]\gd(\gs)\gd'(\bar \gs)
\q\\ \nn
\bar\gb_1&=& \half\big[\gt_1(\gt_1-\gt_2)\gd(\bgt_1)\gd(\bgt_2)\gd(\gs)\gd(\bar \gs)
+( 1- \gt_3\bgt_3)\gt_1 \gd(\bgt_2)\gd(\gs)\gd'(\bar \gs)
\big]\q
\\ \nn
\bar\gb_2&=& \half\big[\gt_2(\gt_2-\gt_1)\gd(\bgt_1)\gd(\bgt_2)\gd(\gs)\gd(\bar \gs)
+( 1- \gt_3\bgt_3)\gt_2 \gd(\bgt_1)\gd (\gs)\gd'(\bar \gs)
\big]\q
\eee
where\be
\gs=  1-\sum_{i=1}^3 \gt_i  \q\bar \gs=  1-\sum_{i=1}^3 \bgt_i \,.\ee
The resulting local vertex is $$G^{loc}|_{\bar H}=-\half \bar H^{\pa\pa}
   \int \int d^3\bgt    d^3\gt \, g_{\pa\pa}^{loc} \exp\big(  \gt_3\p_1{}_{\ga}\p_2{}^{\ga}+\bgt_3\bp_1{}_{\pa}\bp_2{}^{\pa}
  \big)  \PPP(   \gt_1y,  -\gt_2 y ,   \bgt_1\by,-\bgt_2\by;K|x) \,
\qquad
$$ with
 \bee \label{LocRes}
 g^{loc}_{\pa\pa}=      \half\gd(\gt_3) \Big  ( \gd(\gt_1)+ \gd(\gt_2)\Big)\gd(\gs)\gd'(\bar\gs)
\bp_1{}_\pa\bp_2{}_\pa
+        \half\gd(\gt_3)    \Big ( \gd(\bgt_1)+ \gd(\bgt_2)\Big)\gd'(\gs)\gd(\bar\gs)
\bp_1{}_\pa\bp_2{}_\pa\qquad\\
\nn
    - \half
       {\gd(\gt_3)}  \gd (\gs)\gd (\bar \gs)
\big(\gd(\bgt_2)\bp_1{}_\pa +\gd(\bgt_1)\bp_2{}_\pa \big)
\big(\gd(\gt_1)\bp_1{}_\pa +\gd(\gt_2)\bp_2{}_\pa \big) \qquad\\ \nn
+ \f{i}{2}\gd(\gt_3)\gt_2 \big(
 \gd(\gt_1)\gd(\gs)\gd'(\bar \gs) +
 \gd(\bgt_2)\gd'(\gs)\gd (\bar \gs) 
\big)
  \by{}_\pa\bp_1{}_\pa\,
\qquad\\ \nn
    - \f{i}{2}\gd(\bgt_3)  \big[\bgt_1(\bgt_1-\bgt_2)
    \gd(\gt_1)\gd(\gt_2)\gd (\gs)\gd (\bar \gs)
+ \bgt_1 \gd(\gt_2)\gd'(\gs)\gd(\bar \gs) \big]
 \by{}_\pa\bp_1{}_\pa
 \qquad\\\nn
       +  \gd(\bgt_3)
\Big\{
 \half\big[\bgt_1(\gt_1+\gt_3\bgt_2)\gd(\gt_2)\te(\gt_1)
 +\bgt_2(\gt_2+\gt_3\bgt_1)\gd(\gt_1)\te(\gt_2)\big]\gd'(\gs)\gd(\bar \gs)\Big\}
   \by_\pa  \by_\pa     \qquad
\\ \nn -{\gd(\gt_3)}
 \half \gt_2  %
      \gt_1 \Big(\gd(\gt_2)\te(\gt_1)+\gd(\gt_1)\te(\gt_2)\Big)\gd(\gs)\gd'(\bar \gs)
 \by_\pa \by_\pa\qquad\\ \nn
+ \f{i}{2}\gd(\gt_3) \gt_1 \big( \gd(\gt_2)\gd(\gs)\gd'(\bar \gs) +
 \gd(\bgt_1)\gd'(\gs)\gd (\bar \gs) \big)
  \by{}_\pa\bp_2{}_\pa\,
\qquad\\ \nn
    -\f{i}{2} \gd(\bgt_3)   \big[\bgt_2(\bgt_2-\bgt_1)
    \gd(\gt_2)\gd(\gt_1)\gd (\gs)\gd (\bar \gs)
+ \bgt_2 \gd(\gt_1)\gd'(\gs)\gd(\bar \gs) \big]
 \by{}_\pa\bp_2{}_\pa
 \,.\qquad \eee
The complex conjugated case is analogous. This solution is less useful than  Solution II
obtained using  another Ansatz.

 \subsection* {Solution II}

\label{Appendix 3}
Setting $a=b_i=\bar b_i=0$ in \eqref{sourccoef}
as well as $\ga=\gb_i=\bar \gb_i=0$ in \eqref{Firtz}, and considering $ g_{i\,j}$
proportional to $\gd(\gt_1\bgt_1-\gt_2\bgt_2)$ one can make sure that
    \bee\label{g11n}
g_{11}=\Big\{\bgt_1(\bgt_2+\gt_1\bgt_3)\gd(\gs)\gd'(\bar\gs)+\gt_1(\gt_2+\gt_3\bgt_1)\gd'(\gs)\gd(\bar\gs)
-\gt_1\bgt_1 \gd(\gs)\gd(\bar\gs)\Big\}\gd(Z)\Upsilon\,,\quad
\\ \label{g21n}
g_{21}=-\Big\{\bgt_1(\bgt_1+\gt_2\bgt_3)\gd(\gs)\gd'(\bar\gs)+\gt_2(\gt_2+\gt_3\bgt_1)\gd'(\gs)\gd(\bar\gs)
-\gt_2\bgt_1 \gd(\gs)\gd(\bar\gs)\Big\}\gd(Z) \Upsilon\,,\quad
\eee
\bee\label{g22n}
g_{22}=\Big\{\bgt_2(\bgt_1+\gt_2\bgt_3)\gd(\gs)\gd'(\bar\gs)+\gt_2(\gt_1+\gt_3\bgt_2)\gd'(\gs)\gd(\bar\gs)
-\gt_2\bgt_2 \gd(\gs)\gd(\bar\gs)\Big\}\gd(Z)\Upsilon\,,\quad
\\ \label{g12n}
g_{12}=-\Big\{\bgt_2(\bgt_2+\gt_1\bgt_3)\gd(\gs)\gd'(\bar\gs)+\gt_1(\gt_1+\gt_3\bgt_2)\gd'(\gs)\gd(\bar\gs)
-\gt_1\bgt_2 \gd(\gs)\gd(\bar\gs)\Big\}\gd(Z) \Upsilon\,,\quad\\\label{Z}
 Z=\gt_1\bgt_1-\gt_2\bgt_2\q \rule{150pt}{0pt}
\eee
  substituted into $X(\PPP)$ \eqref{sourceredifeta1} solve  equation \eqref{eqdadX}.

Because of the factor of $\gd(Z)$, $\D_{ad}X (\PPP)$
contains   distributions like $\gd(\gt_i )\theta(\gt_i)$ that may be ill defined at $\gt_i=0$.
One can see, however, that by the substitution
 \be\label{Zchange} \gt_1\to(1-\gt_3)\gt_1\q
 \gt_2\to(1-\gt_3)\gt_2\q
 \bgt_1\to(1-\bgt_3)\gt_2\q
 \bgt_2\to(1-\bgt_3)\gt_1\,,\ee
 expression  (\ref{sourceredifeta1})
 acquires a nice form (\ref{sourceredifeta0}) which can be independently checked to solve the problem.

\end{document}